\documentclass[longauth]{aa}  

\usepackage{graphicx}
\usepackage{txfonts}
\usepackage{enumitem}   
\usepackage{multirow}   

\usepackage[colorlinks = true, linkcolor = blue, urlcolor  = blue, citecolor = blue]{hyperref}
\usepackage{euclid}
\usepackage[switch, modulo]{lineno}

\usepackage{xcolor}

\definecolor{red1}{RGB}{180,1,25}
\definecolor{blue1}{RGB}{0,128,255}

\newcommand{\orcid}[1]{} %% define as link to https://orcid.org/#1 if needed

\newcommand{\om}{\Omega_{\rm m}}

\newcommand{\lob}{\lambda^{\rm ob}}
\newcommand{\ltr}{\lambda^{\rm tr}}
\newcommand{\zob}{z^{\rm ob}}
\newcommand{\ztr}{z^{\rm tr}}

\defcitealias{Fumagalli21}{F21}
\defcitealias{Fumagalli-EP27}{EC:F24}

\begin{document} 

    \title{\Euclid: Modelling observational effects in cluster counts and cluster clustering\thanks{This paper is published on behalf of the Euclid Consortium.}}  

%% please do not edit the author list -- contact ECEB Bureau for changes
%%%% Version Thursday 25th of September 2025 12:20:33 PM UT                                             
%%%% Please do not edit the author list -- contact ECEB Bureau for changes
\author{A.~Fumagalli\orcid{0009-0004-0300-2535}\thanks{\email{alessandra.fumagalli@inaf.it}}\inst{\ref{aff1},\ref{aff2}}
\and M.~Costanzi\orcid{0000-0001-8158-1449}\inst{\ref{aff3},\ref{aff1},\ref{aff4}}
\and T.~Castro\orcid{0000-0002-6292-3228}\inst{\ref{aff1},\ref{aff5},\ref{aff4},\ref{aff6}}
\and A.~Saro\orcid{0000-0002-9288-862X}\inst{\ref{aff3},\ref{aff4},\ref{aff1},\ref{aff5},\ref{aff6}}
\and S.~Borgani\orcid{0000-0001-6151-6439}\inst{\ref{aff3},\ref{aff4},\ref{aff1},\ref{aff5},\ref{aff6}}
\and M.~Romanello\orcid{0000-0003-4563-4923}\inst{\ref{aff7},\ref{aff8}}
\and F.~Marulli\orcid{0000-0002-8850-0303}\inst{\ref{aff9},\ref{aff8},\ref{aff10}}
\and E.~Tsaprazi\orcid{0000-0001-5082-4380}\inst{\ref{aff11}}
\and P.~Monaco\orcid{0000-0003-2083-7564}\inst{\ref{aff3},\ref{aff1},\ref{aff5},\ref{aff4},\ref{aff6}}
\and B.~Altieri\orcid{0000-0003-3936-0284}\inst{\ref{aff12}}
\and A.~Amara\inst{\ref{aff13}}
\and L.~Amendola\orcid{0000-0002-0835-233X}\inst{\ref{aff14}}
\and S.~Andreon\orcid{0000-0002-2041-8784}\inst{\ref{aff15}}
\and N.~Auricchio\orcid{0000-0003-4444-8651}\inst{\ref{aff8}}
\and C.~Baccigalupi\orcid{0000-0002-8211-1630}\inst{\ref{aff4},\ref{aff1},\ref{aff5},\ref{aff16}}
\and M.~Baldi\orcid{0000-0003-4145-1943}\inst{\ref{aff7},\ref{aff8},\ref{aff10}}
\and A.~Balestra\orcid{0000-0002-6967-261X}\inst{\ref{aff17}}
\and S.~Bardelli\orcid{0000-0002-8900-0298}\inst{\ref{aff8}}
\and A.~Biviano\orcid{0000-0002-0857-0732}\inst{\ref{aff1},\ref{aff4}}
\and E.~Branchini\orcid{0000-0002-0808-6908}\inst{\ref{aff18},\ref{aff19},\ref{aff15}}
\and M.~Brescia\orcid{0000-0001-9506-5680}\inst{\ref{aff20},\ref{aff21}}
\and S.~Camera\orcid{0000-0003-3399-3574}\inst{\ref{aff22},\ref{aff23},\ref{aff24}}
\and G.~Ca\~nas-Herrera\orcid{0000-0003-2796-2149}\inst{\ref{aff25},\ref{aff26},\ref{aff27}}
\and V.~Capobianco\orcid{0000-0002-3309-7692}\inst{\ref{aff24}}
\and C.~Carbone\orcid{0000-0003-0125-3563}\inst{\ref{aff28}}
\and J.~Carretero\orcid{0000-0002-3130-0204}\inst{\ref{aff29},\ref{aff30}}
\and S.~Casas\orcid{0000-0002-4751-5138}\inst{\ref{aff31}}
\and M.~Castellano\orcid{0000-0001-9875-8263}\inst{\ref{aff32}}
\and G.~Castignani\orcid{0000-0001-6831-0687}\inst{\ref{aff8}}
\and S.~Cavuoti\orcid{0000-0002-3787-4196}\inst{\ref{aff21},\ref{aff33}}
\and K.~C.~Chambers\orcid{0000-0001-6965-7789}\inst{\ref{aff34}}
\and A.~Cimatti\inst{\ref{aff35}}
\and C.~Colodro-Conde\inst{\ref{aff36}}
\and G.~Congedo\orcid{0000-0003-2508-0046}\inst{\ref{aff37}}
\and L.~Conversi\orcid{0000-0002-6710-8476}\inst{\ref{aff38},\ref{aff12}}
\and Y.~Copin\orcid{0000-0002-5317-7518}\inst{\ref{aff39}}
\and F.~Courbin\orcid{0000-0003-0758-6510}\inst{\ref{aff40},\ref{aff41}}
\and H.~M.~Courtois\orcid{0000-0003-0509-1776}\inst{\ref{aff42}}
\and A.~Da~Silva\orcid{0000-0002-6385-1609}\inst{\ref{aff43},\ref{aff44}}
\and H.~Degaudenzi\orcid{0000-0002-5887-6799}\inst{\ref{aff45}}
\and S.~de~la~Torre\inst{\ref{aff46}}
\and G.~De~Lucia\orcid{0000-0002-6220-9104}\inst{\ref{aff1}}
\and A.~M.~Di~Giorgio\orcid{0000-0002-4767-2360}\inst{\ref{aff47}}
\and H.~Dole\orcid{0000-0002-9767-3839}\inst{\ref{aff48}}
\and M.~Douspis\orcid{0000-0003-4203-3954}\inst{\ref{aff48}}
\and F.~Dubath\orcid{0000-0002-6533-2810}\inst{\ref{aff45}}
\and C.~A.~J.~Duncan\orcid{0009-0003-3573-0791}\inst{\ref{aff37},\ref{aff49}}
\and X.~Dupac\inst{\ref{aff12}}
\and S.~Dusini\orcid{0000-0002-1128-0664}\inst{\ref{aff50}}
\and S.~Escoffier\orcid{0000-0002-2847-7498}\inst{\ref{aff51}}
\and M.~Farina\orcid{0000-0002-3089-7846}\inst{\ref{aff47}}
\and R.~Farinelli\inst{\ref{aff8}}
\and F.~Faustini\orcid{0000-0001-6274-5145}\inst{\ref{aff32},\ref{aff52}}
\and S.~Ferriol\inst{\ref{aff39}}
\and F.~Finelli\orcid{0000-0002-6694-3269}\inst{\ref{aff8},\ref{aff53}}
\and P.~Fosalba\orcid{0000-0002-1510-5214}\inst{\ref{aff54},\ref{aff55}}
\and N.~Fourmanoit\orcid{0009-0005-6816-6925}\inst{\ref{aff51}}
\and M.~Frailis\orcid{0000-0002-7400-2135}\inst{\ref{aff1}}
\and E.~Franceschi\orcid{0000-0002-0585-6591}\inst{\ref{aff8}}
\and M.~Fumana\orcid{0000-0001-6787-5950}\inst{\ref{aff28}}
\and S.~Galeotta\orcid{0000-0002-3748-5115}\inst{\ref{aff1}}
\and K.~George\orcid{0000-0002-1734-8455}\inst{\ref{aff56}}
\and B.~Gillis\orcid{0000-0002-4478-1270}\inst{\ref{aff37}}
\and C.~Giocoli\orcid{0000-0002-9590-7961}\inst{\ref{aff8},\ref{aff10}}
\and J.~Gracia-Carpio\inst{\ref{aff57}}
\and A.~Grazian\orcid{0000-0002-5688-0663}\inst{\ref{aff17}}
\and F.~Grupp\inst{\ref{aff57},\ref{aff56}}
\and L.~Guzzo\orcid{0000-0001-8264-5192}\inst{\ref{aff58},\ref{aff15},\ref{aff59}}
\and S.~V.~H.~Haugan\orcid{0000-0001-9648-7260}\inst{\ref{aff60}}
\and W.~Holmes\inst{\ref{aff61}}
\and F.~Hormuth\inst{\ref{aff62}}
\and A.~Hornstrup\orcid{0000-0002-3363-0936}\inst{\ref{aff63},\ref{aff64}}
\and K.~Jahnke\orcid{0000-0003-3804-2137}\inst{\ref{aff65}}
\and M.~Jhabvala\inst{\ref{aff66}}
\and B.~Joachimi\orcid{0000-0001-7494-1303}\inst{\ref{aff67}}
\and E.~Keih\"anen\orcid{0000-0003-1804-7715}\inst{\ref{aff68}}
\and S.~Kermiche\orcid{0000-0002-0302-5735}\inst{\ref{aff51}}
\and A.~Kiessling\orcid{0000-0002-2590-1273}\inst{\ref{aff61}}
\and B.~Kubik\orcid{0009-0006-5823-4880}\inst{\ref{aff39}}
\and M.~K\"ummel\orcid{0000-0003-2791-2117}\inst{\ref{aff56}}
\and M.~Kunz\orcid{0000-0002-3052-7394}\inst{\ref{aff69}}
\and H.~Kurki-Suonio\orcid{0000-0002-4618-3063}\inst{\ref{aff70},\ref{aff71}}
\and A.~M.~C.~Le~Brun\orcid{0000-0002-0936-4594}\inst{\ref{aff72}}
\and S.~Ligori\orcid{0000-0003-4172-4606}\inst{\ref{aff24}}
\and P.~B.~Lilje\orcid{0000-0003-4324-7794}\inst{\ref{aff60}}
\and V.~Lindholm\orcid{0000-0003-2317-5471}\inst{\ref{aff70},\ref{aff71}}
\and I.~Lloro\orcid{0000-0001-5966-1434}\inst{\ref{aff73}}
\and G.~Mainetti\orcid{0000-0003-2384-2377}\inst{\ref{aff74}}
\and D.~Maino\inst{\ref{aff58},\ref{aff28},\ref{aff59}}
\and E.~Maiorano\orcid{0000-0003-2593-4355}\inst{\ref{aff8}}
\and O.~Mansutti\orcid{0000-0001-5758-4658}\inst{\ref{aff1}}
\and O.~Marggraf\orcid{0000-0001-7242-3852}\inst{\ref{aff75}}
\and M.~Martinelli\orcid{0000-0002-6943-7732}\inst{\ref{aff32},\ref{aff76}}
\and N.~Martinet\orcid{0000-0003-2786-7790}\inst{\ref{aff46}}
\and R.~J.~Massey\orcid{0000-0002-6085-3780}\inst{\ref{aff77}}
\and E.~Medinaceli\orcid{0000-0002-4040-7783}\inst{\ref{aff8}}
\and S.~Mei\orcid{0000-0002-2849-559X}\inst{\ref{aff78},\ref{aff79}}
\and Y.~Mellier\inst{\ref{aff80},\ref{aff81}}
\and M.~Meneghetti\orcid{0000-0003-1225-7084}\inst{\ref{aff8},\ref{aff10}}
\and E.~Merlin\orcid{0000-0001-6870-8900}\inst{\ref{aff32}}
\and G.~Meylan\inst{\ref{aff82}}
\and J.~J.~Mohr\orcid{0000-0002-6875-2087}\inst{\ref{aff83}}
\and A.~Mora\orcid{0000-0002-1922-8529}\inst{\ref{aff84}}
\and M.~Moresco\orcid{0000-0002-7616-7136}\inst{\ref{aff9},\ref{aff8}}
\and L.~Moscardini\orcid{0000-0002-3473-6716}\inst{\ref{aff9},\ref{aff8},\ref{aff10}}
\and E.~Munari\orcid{0000-0002-1751-5946}\inst{\ref{aff1},\ref{aff4}}
\and R.~Nakajima\orcid{0009-0009-1213-7040}\inst{\ref{aff75}}
\and C.~Neissner\orcid{0000-0001-8524-4968}\inst{\ref{aff85},\ref{aff30}}
\and S.-M.~Niemi\orcid{0009-0005-0247-0086}\inst{\ref{aff25}}
\and C.~Padilla\orcid{0000-0001-7951-0166}\inst{\ref{aff85}}
\and S.~Paltani\orcid{0000-0002-8108-9179}\inst{\ref{aff45}}
\and F.~Pasian\orcid{0000-0002-4869-3227}\inst{\ref{aff1}}
\and K.~Pedersen\inst{\ref{aff86}}
\and V.~Pettorino\inst{\ref{aff25}}
\and S.~Pires\orcid{0000-0002-0249-2104}\inst{\ref{aff87}}
\and G.~Polenta\orcid{0000-0003-4067-9196}\inst{\ref{aff52}}
\and M.~Poncet\inst{\ref{aff88}}
\and L.~A.~Popa\inst{\ref{aff89}}
\and L.~Pozzetti\orcid{0000-0001-7085-0412}\inst{\ref{aff8}}
\and F.~Raison\orcid{0000-0002-7819-6918}\inst{\ref{aff57}}
\and R.~Rebolo\orcid{0000-0003-3767-7085}\inst{\ref{aff36},\ref{aff90},\ref{aff91}}
\and A.~Renzi\orcid{0000-0001-9856-1970}\inst{\ref{aff92},\ref{aff50}}
\and J.~Rhodes\orcid{0000-0002-4485-8549}\inst{\ref{aff61}}
\and G.~Riccio\inst{\ref{aff21}}
\and E.~Romelli\orcid{0000-0003-3069-9222}\inst{\ref{aff1}}
\and M.~Roncarelli\orcid{0000-0001-9587-7822}\inst{\ref{aff8}}
\and C.~Rosset\orcid{0000-0003-0286-2192}\inst{\ref{aff78}}
\and R.~Saglia\orcid{0000-0003-0378-7032}\inst{\ref{aff56},\ref{aff57}}
\and Z.~Sakr\orcid{0000-0002-4823-3757}\inst{\ref{aff14},\ref{aff93},\ref{aff94}}
\and A.~G.~S\'anchez\orcid{0000-0003-1198-831X}\inst{\ref{aff57}}
\and D.~Sapone\orcid{0000-0001-7089-4503}\inst{\ref{aff95}}
\and B.~Sartoris\orcid{0000-0003-1337-5269}\inst{\ref{aff56},\ref{aff1}}
\and P.~Schneider\orcid{0000-0001-8561-2679}\inst{\ref{aff75}}
\and T.~Schrabback\orcid{0000-0002-6987-7834}\inst{\ref{aff96}}
\and A.~Secroun\orcid{0000-0003-0505-3710}\inst{\ref{aff51}}
\and E.~Sefusatti\orcid{0000-0003-0473-1567}\inst{\ref{aff1},\ref{aff4},\ref{aff5}}
\and G.~Seidel\orcid{0000-0003-2907-353X}\inst{\ref{aff65}}
\and M.~Seiffert\orcid{0000-0002-7536-9393}\inst{\ref{aff61}}
\and S.~Serrano\orcid{0000-0002-0211-2861}\inst{\ref{aff54},\ref{aff97},\ref{aff55}}
\and P.~Simon\inst{\ref{aff75}}
\and C.~Sirignano\orcid{0000-0002-0995-7146}\inst{\ref{aff92},\ref{aff50}}
\and G.~Sirri\orcid{0000-0003-2626-2853}\inst{\ref{aff10}}
\and A.~Spurio~Mancini\orcid{0000-0001-5698-0990}\inst{\ref{aff98}}
\and L.~Stanco\orcid{0000-0002-9706-5104}\inst{\ref{aff50}}
\and J.~Steinwagner\orcid{0000-0001-7443-1047}\inst{\ref{aff57}}
\and P.~Tallada-Cresp\'{i}\orcid{0000-0002-1336-8328}\inst{\ref{aff29},\ref{aff30}}
\and D.~Tavagnacco\orcid{0000-0001-7475-9894}\inst{\ref{aff1}}
\and A.~N.~Taylor\inst{\ref{aff37}}
\and I.~Tereno\orcid{0000-0002-4537-6218}\inst{\ref{aff43},\ref{aff99}}
\and N.~Tessore\orcid{0000-0002-9696-7931}\inst{\ref{aff67}}
\and S.~Toft\orcid{0000-0003-3631-7176}\inst{\ref{aff100},\ref{aff101}}
\and R.~Toledo-Moreo\orcid{0000-0002-2997-4859}\inst{\ref{aff102}}
\and F.~Torradeflot\orcid{0000-0003-1160-1517}\inst{\ref{aff30},\ref{aff29}}
\and I.~Tutusaus\orcid{0000-0002-3199-0399}\inst{\ref{aff55},\ref{aff54},\ref{aff93}}
\and L.~Valenziano\orcid{0000-0002-1170-0104}\inst{\ref{aff8},\ref{aff53}}
\and J.~Valiviita\orcid{0000-0001-6225-3693}\inst{\ref{aff70},\ref{aff71}}
\and T.~Vassallo\orcid{0000-0001-6512-6358}\inst{\ref{aff56},\ref{aff1}}
\and G.~Verdoes~Kleijn\orcid{0000-0001-5803-2580}\inst{\ref{aff103}}
\and A.~Veropalumbo\orcid{0000-0003-2387-1194}\inst{\ref{aff15},\ref{aff19},\ref{aff18}}
\and Y.~Wang\orcid{0000-0002-4749-2984}\inst{\ref{aff104}}
\and J.~Weller\orcid{0000-0002-8282-2010}\inst{\ref{aff56},\ref{aff57}}
\and G.~Zamorani\orcid{0000-0002-2318-301X}\inst{\ref{aff8}}
\and F.~M.~Zerbi\inst{\ref{aff15}}
\and E.~Zucca\orcid{0000-0002-5845-8132}\inst{\ref{aff8}}
\and C.~Burigana\orcid{0000-0002-3005-5796}\inst{\ref{aff105},\ref{aff53}}
\and L.~Gabarra\orcid{0000-0002-8486-8856}\inst{\ref{aff106}}
\and M.~Maturi\orcid{0000-0002-3517-2422}\inst{\ref{aff14},\ref{aff107}}
\and C.~Porciani\orcid{0000-0002-7797-2508}\inst{\ref{aff75}}
\and V.~Scottez\orcid{0009-0008-3864-940X}\inst{\ref{aff80},\ref{aff108}}
\and M.~Sereno\orcid{0000-0003-0302-0325}\inst{\ref{aff8},\ref{aff10}}
\and M.~Viel\orcid{0000-0002-2642-5707}\inst{\ref{aff4},\ref{aff1},\ref{aff16},\ref{aff5},\ref{aff6}}}
                                                                                   
%%%% please do not edit the affiliation list -- contact ECEB Bureau for changes
\institute{INAF-Osservatorio Astronomico di Trieste, Via G. B. Tiepolo 11, 34143 Trieste, Italy\label{aff1}
\and
Ludwig-Maximilians-University, Schellingstrasse 4, 80799 Munich, Germany\label{aff2}
\and
Dipartimento di Fisica - Sezione di Astronomia, Universit\`a di Trieste, Via Tiepolo 11, 34131 Trieste, Italy\label{aff3}
\and
IFPU, Institute for Fundamental Physics of the Universe, via Beirut 2, 34151 Trieste, Italy\label{aff4}
\and
INFN, Sezione di Trieste, Via Valerio 2, 34127 Trieste TS, Italy\label{aff5}
\and
ICSC - Centro Nazionale di Ricerca in High Performance Computing, Big Data e Quantum Computing, Via Magnanelli 2, Bologna, Italy\label{aff6}
\and
Dipartimento di Fisica e Astronomia, Universit\`a di Bologna, Via Gobetti 93/2, 40129 Bologna, Italy\label{aff7}
\and
INAF-Osservatorio di Astrofisica e Scienza dello Spazio di Bologna, Via Piero Gobetti 93/3, 40129 Bologna, Italy\label{aff8}
\and
Dipartimento di Fisica e Astronomia "Augusto Righi" - Alma Mater Studiorum Universit\`a di Bologna, via Piero Gobetti 93/2, 40129 Bologna, Italy\label{aff9}
\and
INFN-Sezione di Bologna, Viale Berti Pichat 6/2, 40127 Bologna, Italy\label{aff10}
\and
Astrophysics Group, Blackett Laboratory, Imperial College London, London SW7 2AZ, UK\label{aff11}
\and
ESAC/ESA, Camino Bajo del Castillo, s/n., Urb. Villafranca del Castillo, 28692 Villanueva de la Ca\~nada, Madrid, Spain\label{aff12}
\and
School of Mathematics and Physics, University of Surrey, Guildford, Surrey, GU2 7XH, UK\label{aff13}
\and
Institut f\"ur Theoretische Physik, University of Heidelberg, Philosophenweg 16, 69120 Heidelberg, Germany\label{aff14}
\and
INAF-Osservatorio Astronomico di Brera, Via Brera 28, 20122 Milano, Italy\label{aff15}
\and
SISSA, International School for Advanced Studies, Via Bonomea 265, 34136 Trieste TS, Italy\label{aff16}
\and
INAF-Osservatorio Astronomico di Padova, Via dell'Osservatorio 5, 35122 Padova, Italy\label{aff17}
\and
Dipartimento di Fisica, Universit\`a di Genova, Via Dodecaneso 33, 16146, Genova, Italy\label{aff18}
\and
INFN-Sezione di Genova, Via Dodecaneso 33, 16146, Genova, Italy\label{aff19}
\and
Department of Physics "E. Pancini", University Federico II, Via Cinthia 6, 80126, Napoli, Italy\label{aff20}
\and
INAF-Osservatorio Astronomico di Capodimonte, Via Moiariello 16, 80131 Napoli, Italy\label{aff21}
\and
Dipartimento di Fisica, Universit\`a degli Studi di Torino, Via P. Giuria 1, 10125 Torino, Italy\label{aff22}
\and
INFN-Sezione di Torino, Via P. Giuria 1, 10125 Torino, Italy\label{aff23}
\and
INAF-Osservatorio Astrofisico di Torino, Via Osservatorio 20, 10025 Pino Torinese (TO), Italy\label{aff24}
\and
European Space Agency/ESTEC, Keplerlaan 1, 2201 AZ Noordwijk, The Netherlands\label{aff25}
\and
Institute Lorentz, Leiden University, Niels Bohrweg 2, 2333 CA Leiden, The Netherlands\label{aff26}
\and
Leiden Observatory, Leiden University, Einsteinweg 55, 2333 CC Leiden, The Netherlands\label{aff27}
\and
INAF-IASF Milano, Via Alfonso Corti 12, 20133 Milano, Italy\label{aff28}
\and
Centro de Investigaciones Energ\'eticas, Medioambientales y Tecnol\'ogicas (CIEMAT), Avenida Complutense 40, 28040 Madrid, Spain\label{aff29}
\and
Port d'Informaci\'{o} Cient\'{i}fica, Campus UAB, C. Albareda s/n, 08193 Bellaterra (Barcelona), Spain\label{aff30}
\and
Institute for Theoretical Particle Physics and Cosmology (TTK), RWTH Aachen University, 52056 Aachen, Germany\label{aff31}
\and
INAF-Osservatorio Astronomico di Roma, Via Frascati 33, 00078 Monteporzio Catone, Italy\label{aff32}
\and
INFN section of Naples, Via Cinthia 6, 80126, Napoli, Italy\label{aff33}
\and
Institute for Astronomy, University of Hawaii, 2680 Woodlawn Drive, Honolulu, HI 96822, USA\label{aff34}
\and
Dipartimento di Fisica e Astronomia "Augusto Righi" - Alma Mater Studiorum Universit\`a di Bologna, Viale Berti Pichat 6/2, 40127 Bologna, Italy\label{aff35}
\and
Instituto de Astrof\'{\i}sica de Canarias, V\'{\i}a L\'actea, 38205 La Laguna, Tenerife, Spain\label{aff36}
\and
Institute for Astronomy, University of Edinburgh, Royal Observatory, Blackford Hill, Edinburgh EH9 3HJ, UK\label{aff37}
\and
European Space Agency/ESRIN, Largo Galileo Galilei 1, 00044 Frascati, Roma, Italy\label{aff38}
\and
Universit\'e Claude Bernard Lyon 1, CNRS/IN2P3, IP2I Lyon, UMR 5822, Villeurbanne, F-69100, France\label{aff39}
\and
Institut de Ci\`{e}ncies del Cosmos (ICCUB), Universitat de Barcelona (IEEC-UB), Mart\'{i} i Franqu\`{e}s 1, 08028 Barcelona, Spain\label{aff40}
\and
Instituci\'o Catalana de Recerca i Estudis Avan\c{c}ats (ICREA), Passeig de Llu\'{\i}s Companys 23, 08010 Barcelona, Spain\label{aff41}
\and
UCB Lyon 1, CNRS/IN2P3, IUF, IP2I Lyon, 4 rue Enrico Fermi, 69622 Villeurbanne, France\label{aff42}
\and
Departamento de F\'isica, Faculdade de Ci\^encias, Universidade de Lisboa, Edif\'icio C8, Campo Grande, PT1749-016 Lisboa, Portugal\label{aff43}
\and
Instituto de Astrof\'isica e Ci\^encias do Espa\c{c}o, Faculdade de Ci\^encias, Universidade de Lisboa, Campo Grande, 1749-016 Lisboa, Portugal\label{aff44}
\and
Department of Astronomy, University of Geneva, ch. d'Ecogia 16, 1290 Versoix, Switzerland\label{aff45}
\and
Aix-Marseille Universit\'e, CNRS, CNES, LAM, Marseille, France\label{aff46}
\and
INAF-Istituto di Astrofisica e Planetologia Spaziali, via del Fosso del Cavaliere, 100, 00100 Roma, Italy\label{aff47}
\and
Universit\'e Paris-Saclay, CNRS, Institut d'astrophysique spatiale, 91405, Orsay, France\label{aff48}
\and
Jodrell Bank Centre for Astrophysics, Department of Physics and Astronomy, University of Manchester, Oxford Road, Manchester M13 9PL, UK\label{aff49}
\and
INFN-Padova, Via Marzolo 8, 35131 Padova, Italy\label{aff50}
\and
Aix-Marseille Universit\'e, CNRS/IN2P3, CPPM, Marseille, France\label{aff51}
\and
Space Science Data Center, Italian Space Agency, via del Politecnico snc, 00133 Roma, Italy\label{aff52}
\and
INFN-Bologna, Via Irnerio 46, 40126 Bologna, Italy\label{aff53}
\and
Institut d'Estudis Espacials de Catalunya (IEEC),  Edifici RDIT, Campus UPC, 08860 Castelldefels, Barcelona, Spain\label{aff54}
\and
Institute of Space Sciences (ICE, CSIC), Campus UAB, Carrer de Can Magrans, s/n, 08193 Barcelona, Spain\label{aff55}
\and
Universit\"ats-Sternwarte M\"unchen, Fakult\"at f\"ur Physik, Ludwig-Maximilians-Universit\"at M\"unchen, Scheinerstrasse 1, 81679 M\"unchen, Germany\label{aff56}
\and
Max Planck Institute for Extraterrestrial Physics, Giessenbachstr. 1, 85748 Garching, Germany\label{aff57}
\and
Dipartimento di Fisica "Aldo Pontremoli", Universit\`a degli Studi di Milano, Via Celoria 16, 20133 Milano, Italy\label{aff58}
\and
INFN-Sezione di Milano, Via Celoria 16, 20133 Milano, Italy\label{aff59}
\and
Institute of Theoretical Astrophysics, University of Oslo, P.O. Box 1029 Blindern, 0315 Oslo, Norway\label{aff60}
\and
Jet Propulsion Laboratory, California Institute of Technology, 4800 Oak Grove Drive, Pasadena, CA, 91109, USA\label{aff61}
\and
Felix Hormuth Engineering, Goethestr. 17, 69181 Leimen, Germany\label{aff62}
\and
Technical University of Denmark, Elektrovej 327, 2800 Kgs. Lyngby, Denmark\label{aff63}
\and
Cosmic Dawn Center (DAWN), Denmark\label{aff64}
\and
Max-Planck-Institut f\"ur Astronomie, K\"onigstuhl 17, 69117 Heidelberg, Germany\label{aff65}
\and
NASA Goddard Space Flight Center, Greenbelt, MD 20771, USA\label{aff66}
\and
Department of Physics and Astronomy, University College London, Gower Street, London WC1E 6BT, UK\label{aff67}
\and
Department of Physics and Helsinki Institute of Physics, Gustaf H\"allstr\"omin katu 2, University of Helsinki, 00014 Helsinki, Finland\label{aff68}
\and
Universit\'e de Gen\`eve, D\'epartement de Physique Th\'eorique and Centre for Astroparticle Physics, 24 quai Ernest-Ansermet, CH-1211 Gen\`eve 4, Switzerland\label{aff69}
\and
Department of Physics, P.O. Box 64, University of Helsinki, 00014 Helsinki, Finland\label{aff70}
\and
Helsinki Institute of Physics, Gustaf H{\"a}llstr{\"o}min katu 2, University of Helsinki, 00014 Helsinki, Finland\label{aff71}
\and
Laboratoire d'etude de l'Univers et des phenomenes eXtremes, Observatoire de Paris, Universit\'e PSL, Sorbonne Universit\'e, CNRS, 92190 Meudon, France\label{aff72}
\and
SKAO, Jodrell Bank, Lower Withington, Macclesfield SK11 9FT, United Kingdom\label{aff73}
\and
Centre de Calcul de l'IN2P3/CNRS, 21 avenue Pierre de Coubertin 69627 Villeurbanne Cedex, France\label{aff74}
\and
Universit\"at Bonn, Argelander-Institut f\"ur Astronomie, Auf dem H\"ugel 71, 53121 Bonn, Germany\label{aff75}
\and
INFN-Sezione di Roma, Piazzale Aldo Moro, 2 - c/o Dipartimento di Fisica, Edificio G. Marconi, 00185 Roma, Italy\label{aff76}
\and
Department of Physics, Institute for Computational Cosmology, Durham University, South Road, Durham, DH1 3LE, UK\label{aff77}
\and
Universit\'e Paris Cit\'e, CNRS, Astroparticule et Cosmologie, 75013 Paris, France\label{aff78}
\and
CNRS-UCB International Research Laboratory, Centre Pierre Bin\'etruy, IRL2007, CPB-IN2P3, Berkeley, USA\label{aff79}
\and
Institut d'Astrophysique de Paris, 98bis Boulevard Arago, 75014, Paris, France\label{aff80}
\and
Institut d'Astrophysique de Paris, UMR 7095, CNRS, and Sorbonne Universit\'e, 98 bis boulevard Arago, 75014 Paris, France\label{aff81}
\and
Institute of Physics, Laboratory of Astrophysics, Ecole Polytechnique F\'ed\'erale de Lausanne (EPFL), Observatoire de Sauverny, 1290 Versoix, Switzerland\label{aff82}
\and
University Observatory, LMU Faculty of Physics, Scheinerstrasse 1, 81679 Munich, Germany\label{aff83}
\and
Telespazio UK S.L. for European Space Agency (ESA), Camino bajo del Castillo, s/n, Urbanizacion Villafranca del Castillo, Villanueva de la Ca\~nada, 28692 Madrid, Spain\label{aff84}
\and
Institut de F\'{i}sica d'Altes Energies (IFAE), The Barcelona Institute of Science and Technology, Campus UAB, 08193 Bellaterra (Barcelona), Spain\label{aff85}
\and
DARK, Niels Bohr Institute, University of Copenhagen, Jagtvej 155, 2200 Copenhagen, Denmark\label{aff86}
\and
Universit\'e Paris-Saclay, Universit\'e Paris Cit\'e, CEA, CNRS, AIM, 91191, Gif-sur-Yvette, France\label{aff87}
\and
Centre National d'Etudes Spatiales -- Centre spatial de Toulouse, 18 avenue Edouard Belin, 31401 Toulouse Cedex 9, France\label{aff88}
\and
Institute of Space Science, Str. Atomistilor, nr. 409 M\u{a}gurele, Ilfov, 077125, Romania\label{aff89}
\and
Consejo Superior de Investigaciones Cientificas, Calle Serrano 117, 28006 Madrid, Spain\label{aff90}
\and
Universidad de La Laguna, Departamento de Astrof\'{\i}sica, 38206 La Laguna, Tenerife, Spain\label{aff91}
\and
Dipartimento di Fisica e Astronomia "G. Galilei", Universit\`a di Padova, Via Marzolo 8, 35131 Padova, Italy\label{aff92}
\and
Institut de Recherche en Astrophysique et Plan\'etologie (IRAP), Universit\'e de Toulouse, CNRS, UPS, CNES, 14 Av. Edouard Belin, 31400 Toulouse, France\label{aff93}
\and
Universit\'e St Joseph; Faculty of Sciences, Beirut, Lebanon\label{aff94}
\and
Departamento de F\'isica, FCFM, Universidad de Chile, Blanco Encalada 2008, Santiago, Chile\label{aff95}
\and
Universit\"at Innsbruck, Institut f\"ur Astro- und Teilchenphysik, Technikerstr. 25/8, 6020 Innsbruck, Austria\label{aff96}
\and
Satlantis, University Science Park, Sede Bld 48940, Leioa-Bilbao, Spain\label{aff97}
\and
Department of Physics, Royal Holloway, University of London, Surrey TW20 0EX, UK\label{aff98}
\and
Instituto de Astrof\'isica e Ci\^encias do Espa\c{c}o, Faculdade de Ci\^encias, Universidade de Lisboa, Tapada da Ajuda, 1349-018 Lisboa, Portugal\label{aff99}
\and
Cosmic Dawn Center (DAWN)\label{aff100}
\and
Niels Bohr Institute, University of Copenhagen, Jagtvej 128, 2200 Copenhagen, Denmark\label{aff101}
\and
Universidad Polit\'ecnica de Cartagena, Departamento de Electr\'onica y Tecnolog\'ia de Computadoras,  Plaza del Hospital 1, 30202 Cartagena, Spain\label{aff102}
\and
Kapteyn Astronomical Institute, University of Groningen, PO Box 800, 9700 AV Groningen, The Netherlands\label{aff103}
\and
Infrared Processing and Analysis Center, California Institute of Technology, Pasadena, CA 91125, USA\label{aff104}
\and
INAF, Istituto di Radioastronomia, Via Piero Gobetti 101, 40129 Bologna, Italy\label{aff105}
\and
Department of Physics, Oxford University, Keble Road, Oxford OX1 3RH, UK\label{aff106}
\and
Zentrum f\"ur Astronomie, Universit\"at Heidelberg, Philosophenweg 12, 69120 Heidelberg, Germany\label{aff107}
\and
ICL, Junia, Universit\'e Catholique de Lille, LITL, 59000 Lille, France\label{aff108}}    

   \date{Received ???; accepted ???}

  \abstract
  % context heading (optional)
  % {} leave it empty if necessary  
   % {}
  % aims heading (mandatory)
   {This study explores the impact of observational and modelling systematic effects on cluster number counts and cluster clustering and provides model prescriptions for their joint analysis in the context of the \Euclid survey.
  % methods heading (mandatory)
   Using 1000 \Euclid-like cluster catalogues, we investigated the effect of systematic uncertainties on cluster summary statistics and their auto- and cross-covariance, and performed a likelihood analysis to evaluate their impact on cosmological constraints, with a focus on the matter density parameter, $\om$, and on the power spectrum amplitude, $\sigma_8$.
  % results heading (mandatory)
   Combining cluster clustering with number counts significantly improves cosmological constraints, with the figure of merit increasing by over 300\% compared to number counts alone. We confirm that the two probes are uncorrelated, and the cosmological constraints derived from their combination are almost insensitive to the cosmology dependence of the covariance. We find that photometric redshift uncertainties broaden cosmological posteriors by 20--30\%, while secondary effects such as redshift-space distortions (RSDs) have a smaller impact on the posteriors -- 5\% for clustering alone, 10\% when combining probes -- but can significantly bias the constraints if neglected. We show that clustering data below $60\,h^{-1}\,$Mpc provides additional constraining power, while scales larger than the acoustic oscillation scale add almost no information on $\om$ and $\sigma_8$ parameters. The RSDs and photo-$z$ uncertainties also influence the number count covariance, with a significant impact of about 15--20\% on the parameter constraints.}
  % conclusions heading (optional), leave it empty if necessary 
   % {}


    \titlerunning{\Euclid: Cluster counts and cluster clustering}
    \authorrunning{A. Fumagalli et al.}
   \maketitle
\section{Introduction} \label{sec:intro}
Galaxy clusters, as tracers of the large-scale structure geometry and evolution, are well-known cosmological probes and are sensitive to the properties of the initial density field of the Universe, to the nature of dark matter and dark energy, and to the laws of gravity on large scales \citep[e.g.][]{Allen:2011zs,Kravtsov:2012zs}. So far, the primary method for extracting cosmological information from cluster catalogues has been through analysis of their mass- and redshift-dependent number counts \citep[e.g.][]{Borgani:2001ir,Henry:2004ig,Vikhlinin:2008ym,Mantz:2014paa,Planck:2015lwi,DES:2020cbm,Lesci:2020qpk,SPT:2024qbr,Ghirardini:2024yni,Lesci:2025bqt}, which allows constraints on the matter density parameter, $\om$, and the amplitude of density fluctuations, $\sigma_8$, in the Universe to be derived. 

Beyond traditional methods such as cluster counts, cluster clustering stands out as a powerful technique for cosmology \citep{Borgani:1998sfa,Estrada:2008em,Marulli:2018owk}. By describing the spatial distribution of galaxy clusters, clustering captures features that are missed by simple cluster abundance, allowing us to deepen our understanding of the Universe's large-scale structure. Despite the limitations of current statistics, which prevent cluster clustering from being a competitive stand-alone probe, this approach has proven to be a valuable tool in breaking degeneracies in cosmological and mass-observable parameters when combined with complementary observables such as cluster abundance \citep{Schuecker:2002ti,Majumdar:2003mw,Mana:2013qba,Lesci:2022owx,Fumagalli:2023yym}. 

The constraining power of number counts and clustering is expected to increase significantly with the upcoming datasets from Stage IV wide-surveys, such as eROSITA\footnote{\url{ http://www.mpe.mpg.de/eROSITA}} \citep{predehl2014}, \textit{Euclid}\,\footnote{\url{http://sci.esa.int/euclid/}} \citep{Laureijs11}, and \textit{Vera C. Rubin} Observatory Legacy Survey of Space and Time\footnote{\url{https://www.lsst.org/}} \citep[LSST;][]{LSSTDarkEnergyScience:2012kar}. In particular, the successful launch of the ESA space mission \Euclid \citep{Scaramella-EP1, EuclidSkyOverview} in July 2023 marks the beginning of its ambitious goal of mapping \num{14000} deg$^2$ of the extragalactic sky and collecting an unprecedented amount of cosmological data. For galaxy clusters, \Euclid is expected to yield a sample of about $10^5$ objects out to redshift $z = 2$ \citep{Sartoris:2015aga}, using photometric and spectroscopic data and through gravitational lensing.
Handling such a large amount of data requires a thorough understanding of the systematic uncertainties, which will be the primary limitation in future cosmological analyses. Even with substantial progress made over the past decade in characterising these systematic effects, attaining the sub-percent accuracy required by \Euclid remains a major challenge.

Accurate cosmological parameter constraints rely on effectively characterising uncertainties through covariance matrices. Direct calculation of these matrices requires computationally expensive simulations, and while approximate methods can help reduce the computational costs \citep{Pope:2007vz,Monaco:2016pys}, they can also yield noisy results unless a large number of mock realisations (on the order of $ 10^3$--$\,10^4$) are generated. Alternatively, covariance can be estimated using data-driven methods such as bootstrap or jackknife, but the results are noisy and can overestimate the covariance, particularly for two-point statistics \citep{Norberg:2008tg}. Analytical methods offer a fast way to compute covariance matrices without heavy computational demands, producing noise-free, cosmology-dependent results \citep[e.g.][]{Scoccimarro:1999kp,Meiksin:1998mu,Hu:2002we,Takada:2013wfa}.  However, accurately describing all contributions remains challenging. This complexity necessitates validation against simulations to determine which contributions are significant at the required statistical precision, sometimes leading to the need for calibration of parameters that cannot be derived analytically \citep{Xu:2012hg,OConnell:2015src,Fumagalli:2022plg}.

\citet[][hereafter \citetalias{Fumagalli21}]{Fumagalli21} and \citet[][hereafter \citetalias{Fumagalli-EP27}]{Fumagalli-EP27} carried out the first steps for the characterisation of theoretical systematics in the cosmological analysis of cluster number counts \citep[see also][]{Payerne:2022alz, Payerne:2024lrv} and cluster clustering by validating (semi-)analytical covariance matrix models for the analysis of the upcoming \Euclid catalogues. This work extends that process by assessing the impact of observational and modelling systematic effects, such as redshift-space distortions in the linear regime \citep[RSDs;][]{Kaiser:1987qv} and photometric redshift (photo-$z$) errors.
We also investigate the possible cross-correlations between the two probes to develop a robust combined likelihood analysis. Another important contribution to the systematics of cluster cosmology arises from projection effects, which induce a correlated scatter between cluster observables and the surrounding large-scale environment. Such effects can bias cluster measurements, modifying the inferred cluster abundance and enhancing the clustering signal \citep[e.g.][]{DES:2018phx,Sunayama:2020wvq}. Since the modelling of projection effects requires a dedicated and detailed treatment, we leave their investigation to future work and neglect them in the present analysis.

This paper is structured as follows. In Sect.~\ref{sec:theory} we present the theoretical framework to model the two statistical probes, that is, cluster counts and the two-point correlation function (2PCF), along with the corresponding covariance models as well as the likelihood models adopted to infer cosmological and mass-observable relation parameters. In Sect.~\ref{sec:data} we describe the simulations analysed in this work and the measurement of the statistics and their numerical covariance matrices. In Sect.~\ref{sec:results} we investigate the presence of cross-correlations (Sect.~\ref{sec:results:crosscov}) and examine the probe combination (Sect.~\ref{sec:results:probes}) as well as the impact of cosmology-dependent covariances (Sect.~\ref{sec:results:cosmocov}). Additionally, we provide a validation of cluster clustering (Sect.~\ref{sec:results:clustering}) and number counts (Sect.~\ref{sec:results:numbercounts}) modelling. Lastly, in Sect.~\ref{sec:conclusions} we discuss our results and summarise our main conclusions.

\section{Theoretical background} \label{sec:theory}
In this section we introduce the theoretical formalism to describe number counts and cluster clustering, along with their covariance matrix models, originally presented in \citetalias{Fumagalli21} and \citetalias{Fumagalli-EP27}, respectively. As demonstrated in~\citet{Fumagalli:2023yym}, number counts and clustering alone do not provide sufficient constraints on scaling relations. As \Euclid has been designed to deliver high quality space-based weak lensing measures, we also incorporate the weak lensing mean mass to further refine the constraints on the scaling relation parameters.
We also describe the likelihood function adopted for the cosmological inference and the summary statistics used to compare the results. Throughout the paper, quantities labelled with `tr' or `ob' denote true intrinsic or observed quantities, respectively, while $P(A|B)$ denotes the conditional probability of $A$ given $B$. Note that we use the Dark Energy survey \citep[DES;][]{DES:2005dhi} as a reference to model certain quantities. While the values adopted for the \Euclid analysis may differ, DES offers realistic benchmarks for this purpose.

\subsection{Number counts} \label{sec:theory:numbercounts}
We describe the expected number counts of galaxy clusters in the $a$-th redshift bin and $i$-th richness bin as follows
\begin{equation}
\label{eq:numbercounts_obs}
     \langle N \rangle_{ai} =  \int_0^\infty \diff \ztr \;\Omega_{\rm sky}\,\frac{\diff V}{\diff z\, \diff \Omega}(\ztr|\Delta \zob_a) \,\langle n(\ztr) \rangle_i\,.         
\end{equation} 
Here, $\Omega_{\rm sky}$ is the survey area in steradians, assuming a homogeneous data quality over the sky area, and the observed comoving volume element per unit redshift and solid angle is given by
\begin{equation} \label{eq:obs_vol}
\frac{\diff V}{\diff z \, \diff \Omega}(\ztr|\Delta \zob_a) =  \int_{\Delta \zob_a} \diff \zob \frac{\diff V}{\diff z \, \diff \Omega}(\ztr)\,P(\zob\,|\,\ztr, \Delta \lob_i)\,,
\end{equation}
where $\Delta \zob_a$ is the observed redshift bin, and the term $P(\zob \,|\, \ztr, \Delta \lob_i)$ is the probability distribution that assigns an observed redshift $\zob$ to a cluster, given the true redshift and the observed richness bin $\Delta \lob_i$. This term incorporates the effect of photometric uncertainty that we assume to be modelled as a Gaussian distribution with a mean equal to $\ztr$ and a scatter
\begin{equation} \label{eq:photoz}
    \sigma_z(\ztr) = \sigma_{z0}(1+\ztr)\,,
\end{equation}
where $\sigma_{z0}$ is the typical photometric redshift uncertainty of the survey associated with the cluster detection error.

The expected number density of haloes in the $i$-th richness bin is given by
\begin{equation}
\label{eq:massfunc_obs}
    \langle n(\ztr) \rangle_i = \int_0^\infty \diff M \; \frac{\diff n}{\diff M}(M, \ztr) \int_{\Delta \lob_i} \diff \lob \, P(\lob \,|\, M, \ztr)\,,
\end{equation}
where $\diff n/ \diff M$ is the halo mass function and $P(\lob \,|\, M, \ztr)$ denotes the observed richness-mass relation
\begin{equation}
\label{eq:mor}
    P(\lob \,|\, M, \ztr) = \int_0^\infty \diff \ltr \, P(\lob \,|\, \ltr, \ztr) \, P(\ltr \,|\, M, \ztr)\,.
\end{equation}
The term  $P(\ltr \,|\, M, \ztr)$ is the intrinsic richness-mass relation. In this work, we assumed a log-normal distribution with mean and scatter given by
\begin{align}
    \langle \ln\ltr | M_{\rm vir} \rangle &= \ln A_\lambda + B_\lambda \ln \left( \frac{M_{\rm vir}}{M_{\rm p}}\right) + C_\lambda \ln \left (\frac{1+z}{1+z_{\rm p}} \right )\,, \label{eq:mor_mean} \\
    \sigma^2_{\ln \ltr|M_{\rm vir}} &= D_\lambda^2 + \frac{\langle \ltr | M_{\rm vir} \rangle - 1}{\langle \ltr | M_{\rm vir} \rangle^2}\,, \label{eq:mor_var}
\end{align}
where $M_{\rm p} = 3\times10^{14}\,h^{-1}\,M_\odot$ and $z_{\rm p} = 0.45$ \citep{DES:2020cbm} are the pivot mass and redshift parameters, while $A_\lambda, B_\lambda, C_\lambda$, and $D_\lambda$ are the mass-observable relation parameters. The variance consists of intrinsic scatter, $D_\lambda$, plus a Poisson contribution.

The second term in Eq.~\eqref{eq:mor}, $P(\lob | \ltr, \ztr)$, represents the observational scatter in the richness-mass relation, caused by photometric noise, background subtraction uncertainties, and projection or percolation effects. We assumed a normal distribution with scatter given by \citep{DES:2020cbm}
\begin{equation} \label{eq:lob_scatter}
    \sigma_{\lambda^{\rm ob}}(\lambda, z) = (0.9 + 0.1 z) \lambda^{0.4}\,.
\end{equation}

For the number count covariance, we followed the model described in \citetalias{Fumagalli21}, adapted to account for richness-mass relation and selection functions. We describe the number count covariance as the sum of shot-noise and sample variance \citep{Hu:2002we},
\begin{equation}
C_{a b i j} = \left \langle N \right \rangle_{a i}  \,\delta_{a b} \;\delta_{i j} +  \left \langle N b \right \rangle_{a i}  \,\left \langle N b \right \rangle_{b j} \,S_{a b} \,,
\label{eq:cov_nc}
\end{equation}
where $\left \langle N \right \rangle_{ai}$ is the prediction for the number counts in the $a$-th redshift bin and $i$-th richness bin (Eq.~\ref{eq:numbercounts_obs}), and $\left \langle Nb \right \rangle_{ai} $ is the prediction for the number counts times the halo bias, computed in an analogous way. The term $S_{ab}$ is the covariance of the linear density field between two redshift bins given by
\begin{equation} \label{eq:nc_sample_cov}
S_{a b} = \int \frac{\diff^3 k}{(2\pi)^3} \;\sqrt{P_{\rm m}(k,\bar{z}_a)\,P_{\rm m}(k,\bar{z}_b)} \,W_{a}(\mathbf{k})  \,W_{b}(\mathbf{k}) \,,
\end{equation}
where $P_{\rm m}(k,\bar{z}_a)$ is the linear matter power spectrum evaluated at the centre of the $a$-th redshift bin, and $W_{a}(\mathbf{k})$ is the window function of the redshift bin. The window function for a redshift slice of a light cone is given in \citet{DES:2018crd} and takes the form
\begin{equation}
W_{a}(\mathbf{k}) = \frac{4\pi}{V_{a}} \int_{\Delta z_{a}} \diff z \;\Omega_{\rm sky}\,\frac{\diff V}{\diff z \,\diff \Omega} \sum_{\ell=0}^{\infty} \sum_{m = -\ell}^{\ell} {\rm i}^\ell \,j_\ell[k\,r(z)] \,Y_{\ell m}(\hat{\mathbf{k}}) \,K_\ell \,,
\end{equation}
where $V_{a}$ is the volume of the redshift slice, $j_\ell[k\,r(z)]$ are the spherical Bessel functions, $Y_{\ell m}(\hat{\mathbf{k}})$ are the spherical harmonics, $\hat{\mathbf{k}}$ is the angular part of the wave-vector, and $K_\ell$ are the coefficients of the harmonic expansion.

\subsection{Two-point correlation function} \label{sec:theory:2pcf}
In this work we describe cluster clustering through the 3D 2PCF monopole. We considered the auto- and cross-correlation function between different richness samples, each one computed in redshift and radial separation bins. 
We modelled the 2PCF in the $a$-th redshift bin, $s$-th radial bin, and $i$-th and $j$-th richness bins as
\begin{equation} \label{eq:2pcf_obs}
    \langle \xi \rangle_{a s i j} = \int \frac{\diff k\,k^2}{2\pi^2} \langle P_{\rm h}(k)\rangle_{a i j}\, \, W_s(k)\,,
\end{equation}
where $W_s(k)$ is the spherical shell window function given by
\begin{equation} \label{eq:shell_window}
     W_s(k) = \int \frac{\diff^3r}{V_s} j_0(kr) = \frac{r_{s,+}^3 W_{\rm th}(k r_{s,+}) - r_{s,-}^3 W_{\rm th}(k r_{s,-})}{r_{s,+}^3 - r_{s,-}^3}\,,
\end{equation}
where $W_{\rm th}(kr)$ is the top-hat window function and $V_s$ is the volume of the $s$-th spherical shell, delimited by $r_{s,-}$ and $r_{s,+}$. 

The term $\langle P_{\rm h}(k)\rangle_{a i j}$ is the halo power spectrum integrated in the $a$-th redshift bin and in the $i$-th and $j$-th richness bins. It can be expressed as the combination of the auto-power spectra in each richness bin as
\begin{equation}
    \langle P_{\rm h}(k)\rangle_{a i j} = \sqrt{\langle P_{\rm h}(k) \rangle_{a i}\,\langle P_{\rm h}(k) \rangle_{a j}}\,,
\end{equation}
where
\begin{equation} \label{eq:pk_a_i}
    \begin{split}
         \langle P_{\rm h}(k) \rangle_{a i}  = \frac{1}{\langle N \rangle_{a i}} \int_0^\infty \diff \ztr& \, \Omega_{\rm sky}\,\frac{\diff V}{\diff \Omega\,\diff z}(\ztr | \Delta \zob_a) \\
     & \times  \langle n(\ztr)\rangle_i \, \langle b_{\rm eff}(\ztr)\rangle_i \,P_{\rm m}(k,\ztr)\,.    
    \end{split}
\end{equation}
In the above equation, $\langle b_{\rm eff}(\ztr)\rangle_i$ is the effective bias for all clusters having an observed richness in the $i$-th richness bin, and it is computed by weighting the halo mass function in Eq.~\eqref{eq:massfunc_obs} by the halo bias.

To take into account the RSD effect and the uncertainty in the photometric redshift measurements, the matter power spectrum in Eq.~\eqref{eq:pk_a_i} had to be modified as \citep{Marulli:2012na, Sereno:2014eea}
\begin{equation}
    P_{\rm m}^{\,\rm obs}(k) = P'(k) + \beta\,P''(k) + \beta^2\,P'''(k)
,\end{equation}
with
\begin{align} 
    P'(k) &= P_{\rm m}(k)\,\frac{\sqrt{\pi}}{2\,k \sigma}\,{\rm erf}(k \sigma)\,, \label{eq:pk_1}\\ 
    P''(k) &= P_{\rm m}(k)\,\frac{\beta}{(k\sigma)^3}\,\bigg[ \frac{\sqrt{\pi}}{2}\,{\rm erf}(k \sigma) - k\sigma\,\exp(-k^2\,\sigma^2)\bigg]\,,  \label{eq:pk_2}\\
    P'''(k) &= P_{\rm m}(k)\,\frac{\beta^2}{(k\sigma)^5}\,\bigg[ \frac{3\sqrt{\pi}}{8}\,{\rm erf}(k \sigma) 
     \notag \\
    &\hspace{2.2cm} - \frac{k\sigma}{4}\,(2k^2\sigma^2 + 3) + \exp(-k^2\,\sigma^2)\bigg]\,, \label{eq:pk_3}
\end{align}
where $\beta = f_{\rm g} / b_{\rm eff}$ is a parameter\,\footnote{Note that $\beta$ adds an extra dependence on redshift and richness in the matter power spectrum, which is to be taken into account in Eq.~\eqref{eq:pk_a_i}.} containing the dependence on the effective bias and the growth factor,  $f_{\rm g} \approx \left[\Omega_{\rm m}(z)\right]^{0.55}$, and $\sigma$ depends on the photo-$z$ uncertainty $\sigma_z$ as
\begin{equation}
    \sigma = \frac{c\,\sigma_z}{H(z)}\,.
\end{equation}

Equations~\eqref{eq:pk_1}--\eqref{eq:pk_3} exclusively incorporate the linear part of the RSDs, commonly referred to as the Kaiser effect \citep{Kaiser:1987qv}. The so-called finger-of-god non-linear distortion introduces perturbations to the power spectrum similar in shape to those caused by photometric redshift uncertainties. However, its impact is significantly smaller compared to that of photo-$z$ uncertainties and can therefore be safely neglected \citep{Veropalumbo:2013cua,Sereno:2014eea,Romanello:2024kbo}.
The Kaiser effect produces a scale-independent boost of the 2PCF, while the effect of the photo-$z$ uncertainties is to decrease the correlation on small scales and increase it on larger scales, smearing the baryon acoustic oscillation (BAO) peak. 

The BAO scales are also subject to broadening and  a shift of the peak due to non-linear damping, which can be corrected through the infrared resummation \citep[IR;][]{Senatore:2014via,Baldauf:2015xfa}. At lowest order, the matter power spectrum is  corrected as
\begin{equation}
    P_{\rm m}(k,z) \simeq P_{\rm nw}(k,z)+{\rm e}^{-k^2\Sigma^2(z)}P_{\rm w}(k,z)\,,
\end{equation}
where $P_{\rm w}$ and $P_{\rm nw}$ are the wiggled and smooth parts of the linear power spectrum, respectively, and
\begin{equation}
        \Sigma^2(z) = \int_0^{k_s}\frac{\diff q}{6\pi^2}\, P_{\rm nw}(q,z)\,\left[1-j_0\left(q\,r_{\rm BAO}\right)+2j_2\left(q\,r_{\rm BAO}\right)\right]\,.
\end{equation}

Lastly, when measuring the 2PCF from data, it is necessary to make assumptions about the underlying cosmology to convert redshifts into distances. To take this assumption into account, a parameter was introduced to model the geometric distortions (GDs) in the 2PCF shape \citep{Marulli:2012na,Marulli:2015jil}:
\begin{equation}
    \xi(r) \longrightarrow \xi(\alpha r)\,
,\end{equation}
where $r$ is the radial separation and
\begin{equation}
    \alpha = \frac{D_{\rm V}}{r_{\rm s}} \frac{r_{\rm s}^{\rm fid}}{D_{\rm V}^{\rm fid}}\,.
\end{equation}
Here, $r_{\rm s}$ is the position of the sound horizon at decoupling, $D_{\rm V}$ is the isotropic volume distance, and the label `fid' indicates the quantities evaluated at the fiducial cosmology assumed in the measurement process.

We describe the clustering covariance with the semi-analytical model presented in \citetalias{Fumagalli-EP27}. The cross-covariance between different richness samples $i$, $j$, $k$, and $l$, in the $a$-th redshift bin and between the $s$-th and $r$-th separation bins is given by
\begin{equation}
    \label{eq:cov_cl}
    \begin{split}
     C_{asr}^{ijkl}& =  \frac{1}{V_a} \int_0^\infty \frac{\diff k\,k^2}{2 \pi^2}  \left[ \langle P_{\rm h}(k) \rangle_{a i k} + \frac{\delta_{ik}}{\langle n \rangle_{a i}} \right ] \\
     & \hspace{1.8cm} \times  \left[ \langle P_{\rm h}(k)\rangle_{a j l} + \frac{\delta_{jl}}{\langle n \rangle_{a j}} \right] \,W_s(k)\,W_r(k)\\
    & +  \frac{1}{V_a} \int_0^\infty \frac{\diff k\,k^2}{2 \pi^2}\, \langle P_{\rm h}(k)\rangle_{a i j} \, \frac{\delta_{i k}}{\langle n \rangle_{a i}}\, \frac{\delta_{jl}}{\langle n \rangle_{a j}}\frac{W_r(k)}{V_s} \delta_{sr}\\
    & + (k \longleftrightarrow l)\,, 
    \end{split}
\end{equation}
where the term $(k \longleftrightarrow l)$ denotes that the preceding expression should be repeated with the indices $k$ and $l$ swapped.
The first integral represents the standard Gaussian covariance, while the second one is the low-order non-Gaussian contribution.

According to \citetalias{Fumagalli-EP27}, the bias and shot-noise terms in the clustering covariance must be modified by the addition of three parameters $\{\alpha_{\rm c},\,\beta_{\rm c},\,\gamma_{\rm c}\}$ such that
\begin{align*}
   b &\longrightarrow \beta_{\rm c} \, b\,,\\
   \frac{1}{n} &\longrightarrow \frac{1+\alpha_{\rm c}}{n} \ \text{in the Gaussian term}\,,\\
   \frac{1}{n} &\longrightarrow \frac{1+\gamma_{\rm c}}{n} \ \text{in the non-Gaussian term}.
\end{align*}
These parameters can be fitted from a few simulations \citep[of the order of $10^2$;][]{Fumagalli:2022plg} in each redshift and richness bin; the reference values are $\alpha_{\rm c}=0$, $\beta_{\rm c}=1$, and $\gamma_{\rm c}=0$. As shown in \citet[][their figure 8]{Fumagalli:2025twg}, the cosmology dependence of such parameters is absent for $\beta_{\rm c}$ and $\gamma_{\rm c}$, and it is almost negligible for $\alpha_{\rm c}$, thus allowing us to perform a single calibration at the fiducial cosmology of the simulations (Sect.~\ref{sec:data:sims}).

\subsection{Weak lensing mean mass}
Similar to the number counts approach, the expected value of the mean cluster mass within the $a$-th redshift bin and $i$-th richness bin is given by
\begin{equation}
    \label{eq:mean_mass_wl}
    \overline{M}_{a i} = \frac{\langle M^{\rm tot}(\Delta \lambda_i, \Delta z_a) \rangle}{\langle N(\Delta \lambda_i, \Delta z_a) \rangle}\,,
\end{equation}
where $\langle M^{\rm tot} \rangle$ represents the total mass associated with clusters identified within a given redshift and richness range, expressed as
\begin{equation}
\label{eq:mtot_obs}
     \langle M^{\rm tot}(\Delta z_i, \Delta \lambda_j) \rangle =  \int_0^\infty \mathrm{d} \ztr \, \Omega_{\rm sky}\frac{\mathrm{d} V}{\mathrm{d} z \, \mathrm{d} \Omega}(\ztr | \Delta \zob_a) \, \langle M\,n(\ztr)\rangle_i\,,    
\end{equation} 
where the term $\langle M\,n(\ztr) \rangle_i$ is computed by weighting the halo mass function in Eq.~\eqref{eq:massfunc_obs} by the halo mass and accounts for the expected mass contribution from clusters at a given redshift and richness interval.

\subsection{Likelihood function} \label{sec:theory:logl}
We evaluated the impact of systematic effects on cosmological constraints by performing a Bayesian inference on \Euclid-like cluster catalogues derived from 1000 simulated light cones sharing the same underlying cosmological model (see Sect.~\ref{sec:data:sims}). We explored the posterior distribution by using the python wrapper for the nested sampling \texttt{PolyChord} \citep{Handley:2015fda}. 

For number counts and clustering, as well as for weak lensing log-masses, we adopted Gaussian likelihoods,
\begin{equation} \label{eq:gauss_logl}
    \mathcal{L}(\mathbf{x}\,\vert\,\boldsymbol{\mu},\, \tens{C}) = \frac{ \exp {\left \{-\frac{1}{2} (\mathbf{x}-\boldsymbol{\mu})^T C^{-1}  (\mathbf{x}-\boldsymbol{\mu}) \right \}}}{\sqrt{2\pi \det[ \tens{C}]}} \,,
\end{equation}
where $\mathbf{x}$ and $\boldsymbol{\mu}$ are, respectively, the measurement and the model prediction, and $\mathbf{C}$ is the covariance matrix.
We assumed weak lensing masses to be uncorrelated with both counts and clustering based on the assumptions made in \citet{DES:2018crd} and \citet{Fumagalli:2023yym}.  We also considered number counts and cluster clustering as independent probes, following \citet{Mana:2013qba} and \citet{Fumagalli:2023yym}, and we further test the absence of cross-correlation through the use of simulations (see Sect.~\ref{sec:results:crosscov}). Unless differently specified, for cluster counts and clustering, we used cosmology-dependent covariances computed through the analytical models of Eqs.~\eqref{eq:cov_nc} and~\eqref{eq:cov_cl}, respectively. 

Both the cosmological parameters and the parameters describing the mass-observable relation are determined by maximising the average log-likelihood computed across all the $N_{\rm s}$ simulated catalogues; this approach helps mitigate the cosmic variance effect characterising each Universe realisation by a factor of $\sqrt{N_{\rm s}}$. As a result, we are able to detect potential systematic effects in the analysis, identified by any residual shift of the posteriors with respect to the input cosmology.

To assess the difference in the final cosmological posteriors, we use two summary statistics: the figure of merit \citep[FoM;][]{Albrecht:2006um}, which measures the relative amplitude of the posteriors compared to a reference case; and the index of inconsistency \citep[IoI;][]{Lin:2017ikq}, which quantifies their shift from the fiducial cosmology. By running multiple realisations of the same chain, we find that the statistical noise associated with the convergence of likelihood maximisation produces an error of less than 4\% on the FoM and 1\% on the IoI.

\section{Simulated data} \label{sec:data}
In this section we describe the simulated light cones adopted in this work. We also outline the procedure for the measuring of numerical covariances.

\begin{figure*}[t]
    \centering
    \includegraphics[width=0.49\textwidth]{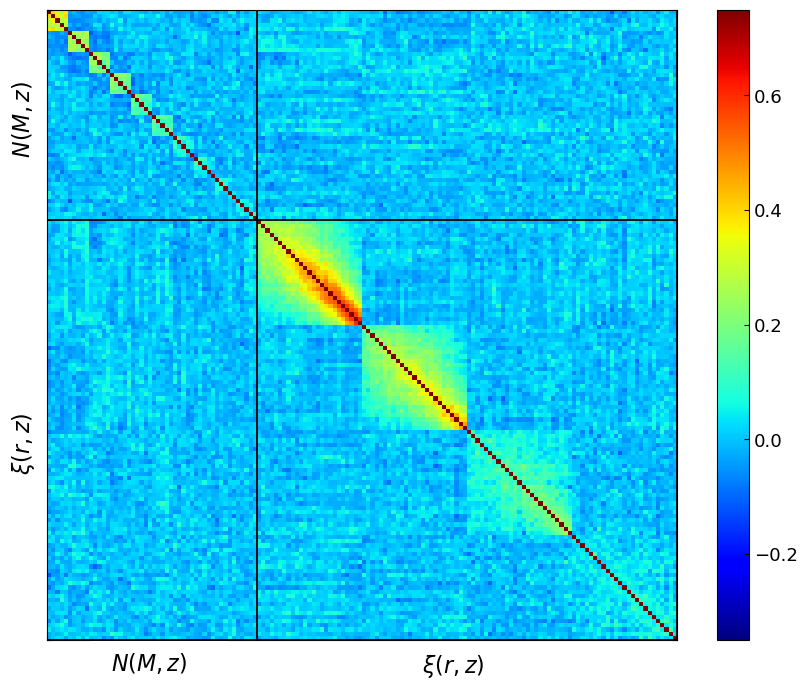}
    \includegraphics[width=0.43\textwidth]{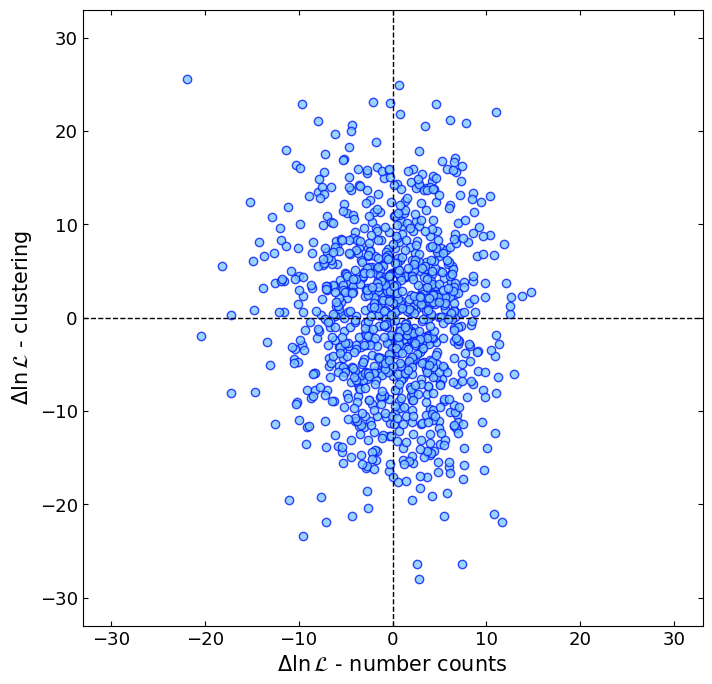}
    \caption{Cross-covariance between number counts and clustering. For better visualisation, we used redshift bins of width $\Delta z=0.2$ for counts and $\Delta z=0.5$ and a single richness bin for the 2PCF. \textit{Left}: Auto- and cross-correlation matrix of number counts and clustering, computed from 1000 mocks. \textit{Right}: Log-likelihood residuals for number counts and clustering for each one of the 1000 light cones, with respect to the mean value assuming the fiducial model parameters.}
    \label{fig:cross_cov}
\end{figure*}

\subsection{Simulations} \label{sec:data:sims}
Accurate covariance matrix estimation necessitates extensive datasets, often exceeding $10^3$ catalogues \citep{Taylor:2012kz,Dodelson:2013uaa}. Instead of computationally expensive $N$-body simulations, we employed the \texttt{PINOCCHIO} code \citep{Monaco:2001jg,Munari:2016aut}, which utilises third-order Lagrangian perturbation theory and ellipsoidal collapse to quickly generate dark matter halo catalogues.
\texttt{PINOCCHIO} generates initial density fields on a regular grid, and uses Lagrangian perturbation theory to first compute particle collapse times, and then to identify haloes and trace their evolution to final positions. Large simulation boxes were used to construct past light cones, capturing haloes causally connected to an observer at present time. 
As shown in \citet[][their Sect. 4.2]{Fumagalli:2025twg}, on linear scales \texttt{PINOCCHIO} produces numerical covariances that accurately match those obtained from $N$-body simulations.

Our dataset consists of 1000 light cones covering 10\,313 deg$^2$ and redshifts $z=0$ to $z=2.5$. These catalogues, generated from boxes of side $3870\,h^{-1}$\,Mpc and $2160^3$ particles, contain haloes with $M_{\rm vir} \ge 10^{13.4}\,h^{-1}\,M_\odot$. The simulations are based on a flat $\Lambda$CDM cosmology consistent with \citet{Planck:2013pxb}: $\Omega_{\rm m } = 0.30711$, $\Omega_{\rm b} = 0.048254$, $h = 0.6777$, $n_{\rm s} = 0.96$, $\sum m_\nu = 0.06\,{\rm eV}$, $A_{\rm s} = 2.21 \times 10^{-9}$, and $\sigma_8 = 0.8288$.
Similarly to \citetalias{Fumagalli21}, we rescaled the halo masses to match, on average, a reference halo mass function model, while preserving shot-noise and sample variance fluctuations in each mock. In this work we assume the \citet{Castro-EP24} halo mass function. We describe the halo bias with the \citet{EP-Castro24} model, computed and calibrated for the assumed halo mass function. Both models have been calibrated to a percent level of accuracy, ensuring that any uncertainty in the formalism used for modelling these two quantities is negligible. The resulting catalogues contain around $ 10^5$ haloes with $M_{\rm vir} \geq 10^{13.5} h^{-1}\,M_\odot$ each. Each object is labelled with a rescaled virial mass, a true undistorted redshift, and a redshift with RSDs, along with angular coordinates.

To mimic the observational properties of cluster surveys, we attached a true richness to each mass by assuming a scaling relation with mean and variance from Eqs.~\eqref{eq:mor_mean} and ~\eqref{eq:mor_var}, with parameters $A_\lambda = 50$, $B_\lambda = 1.0$, $C_\lambda = 0.01$, and  $D_\lambda = 0.2$, obtained from \citet{DES:2020cbm}.\footnote{The parameter values are converted from $\Delta_{500}$ with respect to the critical density to the virial overdensity.} Such a richness is scattered to produce an observed richness by assuming a log-normal distribution with scatter from Eq.~\eqref{eq:lob_scatter}.

In addition, two observed redshifts are assigned assuming a normal distribution with scatter given by Eq.~\eqref{eq:photoz}: for the typical photo-$z$ uncertainty, we consider an optimistic ($\sigma_0 = 0.005$) and a more realistic ($\sigma_0 = 0.01$) estimate of the expected \Euclid photo-$z$ errors \citep{Q1-SP050}. We define the values based on the photo-$z$ uncertainties of Stage-III surveys \citep[see, for instance,][]{Maturi:2018fge,DES:2020ahh,DES:2025xii}, which serve as a benchmark that Stage-IV surveys aim to improve.
Unless otherwise specified, we assume the conservative scenario as the default in the analysis.

Apart from introducing scatter around the true value, photo-$z$ measurements can also bias cluster redshifts. However, since the \textit{Euclid} requirement of $\Delta z = 0.002$ \citep{Laureijs11} results in a negligible impact on our analysis (see Appendix~\ref{appA}), we did not apply any shift to our redshifts.

Lastly, it should be noted that while the catalogues effectively replicate observational conditions, this work does not account for the effects of any survey mask or inhomogeneous sky coverage, and we assume a simplified model of $P(\lob|\ltr,\ztr)$ that does not take into account the correlation with the LSS, since this function has not yet been calibrated for the \Euclid data.

\subsection{Measurements of the summary statistics} \label{sec:data:measure}
We measure the number counts assuming twenty redshift bins of width $\Delta z = 0.1$ in the range $z = 0$--$2$, and five log-spaced richness bins in the range $\lambda = 20$--$400$. 

The same binning scheme is adopted for weak lensing masses. However, since our simulations do not include lensing information, we generate synthetic measurements of the mean weak lensing mass. This is done by predicting a true mean mass in each richness and redshift bin and perturbing it with a Gaussian distribution, assuming a diagonal constant error of 1\,\%, which aligns with the precision target for Stage IV surveys.\footnote{Since weak lensing masses are used here only as a supporting probe and are not the focus of this study, investigating their measurement uncertainties falls outside the scope of this paper.} To match the number of simulations,  we generate 1000 sets of synthetic measurements by varying the seed of the Gaussian scatter.
According to \citet{DES:2018crd}, we model the cosmological dependence on $\om$ of the weak lensing mass estimates as
\begin{equation} \label{eq:wl_mass}
    \log_{10} \frac{\hat{M}_{\rm WL}}{M_\odot}(\om) = \log_{10} \frac{\hat{M}_{\rm WL}}{M_\odot} \big \vert_{\om=0.3} + \frac{\diff \log_{10}\frac{M_{\rm WL}}{M_\odot}}{\diff \om} (\om-0.3)\,,
\end{equation}
where the slopes derived in each bin from fitting such an equation to the data are listed in Table 1 of \citet{DES:2018crd}, and are used in our cosmological analysis to re-scale the `observed' mean mass at each step of the MCMC. 

We measured the 2PCF by using the \citet{Landy:1993yu} estimator, implemented via the \texttt{CosmoBolognaLib} package \citep{Marulli:2015jil}. To construct the random catalogue, we randomly extracted a subset of objects from each mock, shuffled their coordinates, and combined them into a single catalogue containing $N_{\rm R} = 20\,N_{\rm D}$ objects randomly distributed within the light cone volume.
We assumed five redshift bins of width $\Delta z = 0.4$ and two richness bins defined as $\lambda = [20, 40, 400]$. For the standard analysis, we used 25 separation bins in the range $r = 20$--$130\,h^{-1}\,$Mpc.

\section{Results and Discussion} \label{sec:results}
In this section we present the results of the covariance and joint analysis validation. 
Our analysis begins with an assessment of potential cross-correlations between the two probes, that is, number counts and 2PCF monopole, using numerical simulations. We then compare the cosmological posteriors from different probe combinations to evaluate their constraining power for a \Euclid-like survey.  Next, we examine the impact of photo-$z$ uncertainties and redshift distortions on the clustering and number counts modelling, and we assess the impact on the cosmological constraints obtained through the likelihood analysis of clustering alone, and in combination with number counts. In both cases, we add the weak lensing mass likelihood to break the degeneracies between cosmological and scaling-relation parameters.  
For the likelihood analysis, we assume flat priors on both the cosmological parameters $\om$ and $\logten A_{\rm s}$, and on the four scaling-relation parameters of Eqs.~\eqref{eq:mor_mean} and ~\eqref{eq:mor_var}. Specifically, we use $\om \in [0.2, 0.4]$, $\logten A_{\rm s} \in [-9.0, -8.5]$, $A_\lambda \in [20,80]$, $B_\lambda \in [0.5,1.5]$, $C_\lambda \in [-0.5,0.5]$, and $D_\lambda \in [0.01,0.5]$. All the other parameters are fixed to the fiducial values, while $\sigma_8$ is a derived parameter.

\subsection{Cross-covariance} \label{sec:results:crosscov}
To properly combine cluster counts and cluster clustering in the likelihood analysis, we should account for the cross-correlation between the two probes. We compute the auto- and cross-covariance matrices for the two probes from simulations, as shown in the left panel of Fig.~\ref{fig:cross_cov}. The cross-correlation term is fully dominated by noise and consistent with zero. The result holds independently of redshift. 
As a further proof, we compute the number counts and clustering likelihoods for each light cone, and in the right panel of Fig.~\ref{fig:cross_cov} we show their difference with respect to the log-likelihood averaged over the entire sample of simulations. We quantify the correlation by computing the Pearson correlation coefficient between the log-likelihood residuals. For this purpose, we assumed that the residuals are Gaussian distributed around zero, with a covariance given by
\begin{equation}
    C = 
\begin{pmatrix}
\sigma_{\rm NC}^2 & \rho\,\sigma_{\rm NC}\,\sigma_{\rm CL}\\
\,\rho\,\sigma_{\rm NC}\,\sigma_{\rm CL}& \sigma_{\rm CL}^2
\end{pmatrix}\,,
\end{equation}
where $\sigma_{\rm NC}$ and $\sigma_{\rm CL}$ are, respectively, the standard deviations computed from the number counts and clustering residuals, and $\rho$ is the correlation coefficient.  
We computed the Pearson coefficient from the data, obtaining $\rho = -0.011 \pm 0.031$, where the error was estimated using bootstrap. The absence of correlation in the distribution of points confirms the absence of correlation between the two observables. 
According to the result, number counts and clustering can be considered as two independent probes.
The independence of number counts and clustering is due to the fact that the former is a measure of the average comoving number density of tracers, while the latter measures the fluctuations around this mean value. Since fluctuations around the mean can be arbitrarily varied within a sufficiently large volume of the Universe without altering the mean density of tracers, the two probes turn out to be independent of each other.

\subsection{Probe combination} \label{sec:results:probes}
\begin{figure}[t]
    \centering
    \includegraphics[width=0.49\textwidth]{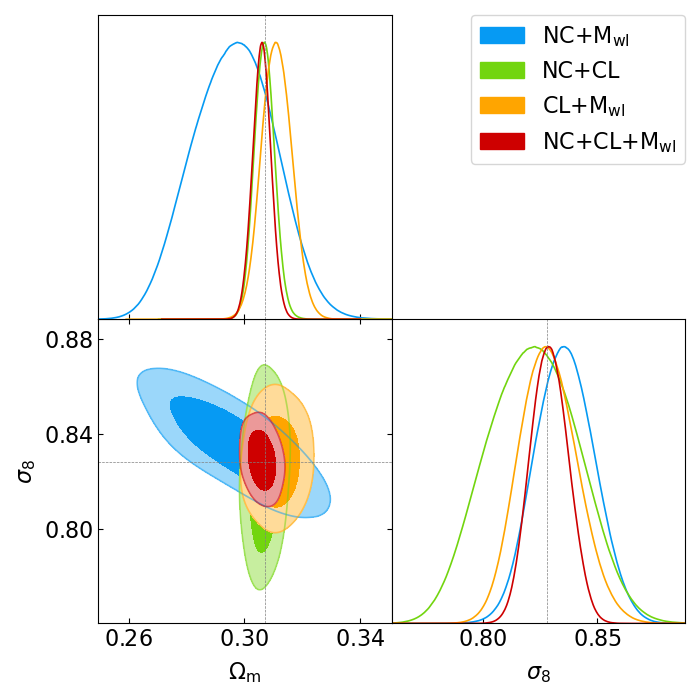}
    \caption{Parameter posterior distributions with 68\% and 95\% confidence intervals from different combinations of probes: number counts and weak lensing mass (in blue), clustering and weak lensing mass (in orange), number counts and clustering (in green), and all the three probes (in red). Dotted grey lines are the input cosmology of the catalogues. The 1D marginalized distributions on the diagonal blocks are normalized to the peak probability. }
    \label{fig:probes_post}
\end{figure}

We explore different combinations of the three probes -- number counts, 2PCF monopole, and weak lensing mean mass -- to evaluate their constraining power for a \Euclid-like survey. As shown in Fig.~\ref{fig:probes_post}, our results align with those of \citet{Fumagalli:2023yym}, confirming the benefits of incorporating cluster clustering as a cosmological probe.

Examining the number counts and weak lensing mass posteriors (blue contours), we note a slight bias, though at a low significance (${\rm IoI} = 0.24$), relative to the fiducial cosmology. This could stem from prior-volume effects. Smaller biases are also visible in other two-probe combinations. However, the full-probe combination is correctly centred, demonstrating that integrating the three probes produces not only more precise, but also more accurate results. This combination yields the tightest constraints, with clustering and weak lensing mass providing the second strongest constraints (see Table~\ref{table:fom_probes}). Therefore, in the subsequent analysis, we primarily focus on these two cases.

\begin{table}[t]
\centering           
\caption{Summary statistics for the impact of different probe combinations on $\om$--$\sigma_8$ posteriors.}
\centering
\begin{tabular}{l c c}  
\hline  
 & $\Delta {\rm FoM} / {\rm FoM}$ & IoI \\
\hline
NC+M$_{\rm WL}$     &   --         & $+\,0.24$  \\
CL+M$_{\rm WL}$     & $-\,74$\,\%  & $+\,0.11$  \\
NC+CL               & $+\,62$\,\%  & $+\,0.02$  \\
NC+CL+M$_{\rm WL}$  & $+\,334$\,\% & $+\,0.01$  \\
\hline
\end{tabular}
\tablebib{FoM residuals are computed with respect to the number counts and weak lensing mass combination. IoI is computed with respect to the fiducial cosmology and scaling relation parameters.}
\label{table:fom_probes} 
\end{table}

\begin{figure}[t]
    \centering
    \includegraphics[width=0.49\textwidth]{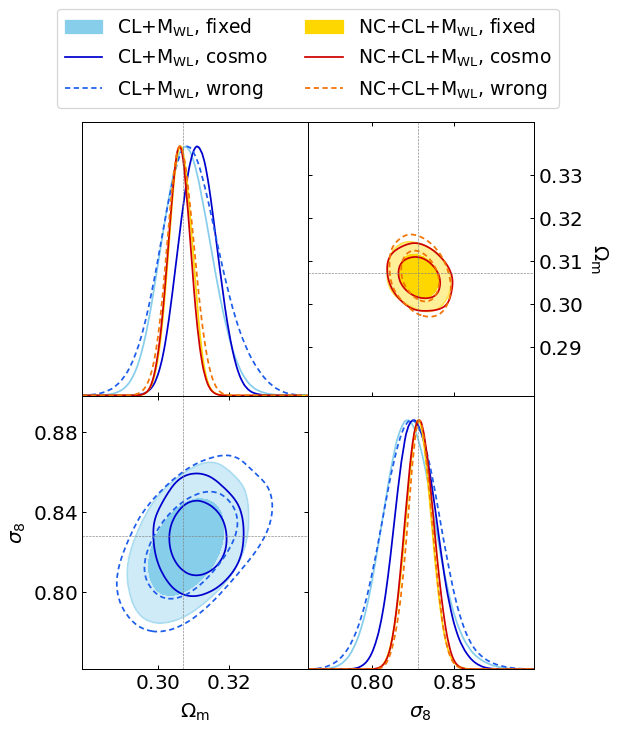}
    \caption{Marginalised posterior distributions in the $\om$--$\sigma_8$ plane, with 68\% and 95\% confidence intervals, when using a fixed-cosmology (filled contours), a cosmology-dependent (empty contours), or the wrong ($\om=0.320$ and $\sigma_8=0.775$, dashed empty contours) covariance matrix. \textit{Lower left}: Clustering and weak lensing masses alone. \textit{Upper right}: Full-probe combination. }
    \label{fig:cosmo_cov}
\end{figure}

\subsection{Cosmology dependence of the covariances}  \label{sec:results:cosmocov}
Here, we assess the impact of cosmology-dependent covariances on cosmological posteriors. As demonstrated in \citetalias{Fumagalli21} and \citetalias{Fumagalli-EP27}, fixing the covariance matrix at an incorrect cosmology can alter the amplitude of the posteriors,  whereas using a cosmology-dependent covariance matrix avoids this issue. Additionally, allowing the clustering covariance to vary freely in the likelihood analysis can provide extra cosmological information. As shown in the lower-left panel of Fig.~\ref{fig:cosmo_cov}, we confirm that despite the broader contours caused by uncertainties in the scaling relation parameters, the clustering covariance still improves the cosmological constraints from cluster clustering, leading to a $40\%$ increase in the FoM. Also, the use of a fixed covariance computed with a wrong cosmology can significantly affect the posteriors' amplitude. As an example, a covariance evaluated at $\om = 0.32$ and $\sigma_8 = 0.775$ produces a $50\%$ difference in the FoM with respect to the input-cosmology case, despite the small parameter shifts: $\Delta\om = 0.013$ and $\Delta\sigma_8= 0.053$.

On the other hand, the impact of the cosmological dependence of the covariance decreases when combining clustering and counts (upper right panel). The improvement observed by using the cosmology-dependent clustering covariance disappears when incorporating the number counts likelihood since the latter already extracts all the additional information encoded in the clustering covariance. This confirms that there is no double counting of the halo mass function information. If such an effect were present, we would expect an additional tightening of the contours when combining number counts with a cosmology-dependent clustering covariance, compared to the fixed-covariance case. However, this is not observed. Further discussion on this can be found in \citetalias{Fumagalli-EP27} and \citet{Fumagalli:2023yym}. Lastly, fixing the covariance matrix to the same wrong cosmology only produces a 6\% difference in the FoM. 
In the following, we always use cosmology-dependent covariances.

\subsection{Cluster clustering validation} \label{sec:results:clustering}
In this section we examine the impact of photo-$z$ distortions (Sect.~\ref{sec:results:clustering:phz}), and secondary distortions (Sect.~\ref{sec:results:clustering:zdist}) on the 2PCF monopole, as well as on its covariance matrix. 
In Sect.~\ref{sec:results:clustering:scales}, we assess how the choice of separation range in the 2PCF measurement influences the constraining power in the cosmological analysis.

\subsubsection{Impact of photo-$z$} \label{sec:results:clustering:phz}
\begin{figure*}[t]
    \centering
    \includegraphics[width=0.5\textwidth]{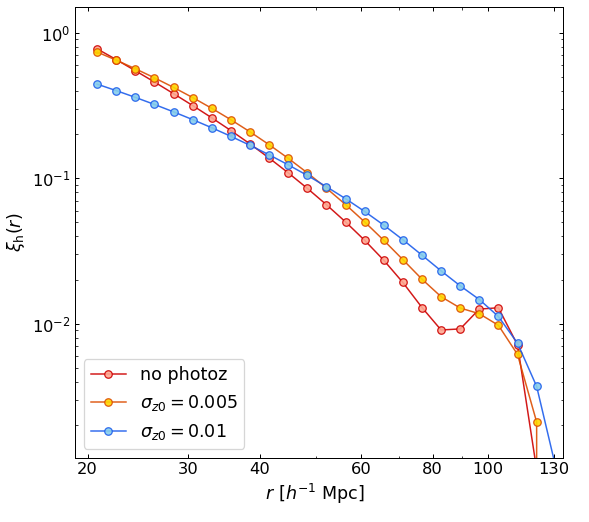}
    \includegraphics[width=0.45\textwidth]{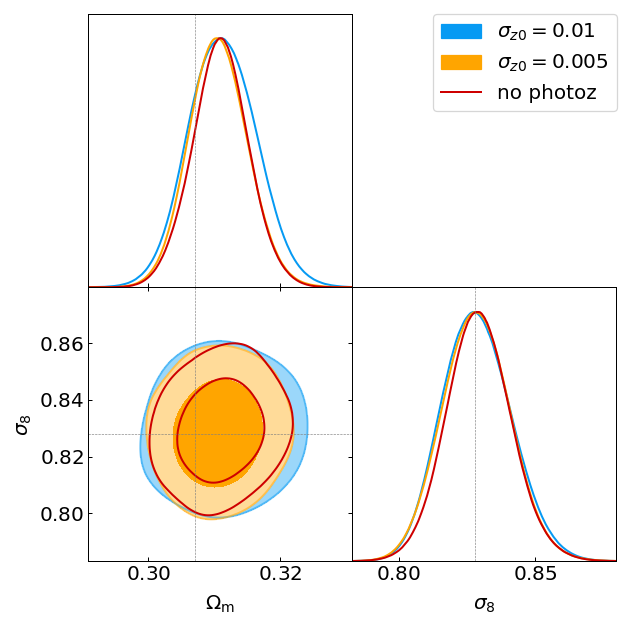}
    \caption{Impact of photo-$z$ uncertainties on the 2PCF. \textit{Left}: Prediction of the 2PCF for different values of photo-$z$ uncertainty. The values are indicated as follows: $\sigma_{z0}=0$ in red, $\sigma_{z0}=0.005$ in orange, and $\sigma_{z0}=0.01$ in blue. This is for the redshift bin $z = 0.4$--$0.8$. \textit{Right}: Marginalised posterior distributions in the $\om$--$\sigma_8$ plane, with 68\% and 95\% confidence intervals, from the combination of clustering and weak lensing masses for the three cases in the left panel.}
    \label{fig:post_photoz}
\end{figure*}

In the case of a photometric survey, the primary dilution in the 2PCF arises from photo-$z$ uncertainties. As shown in the left panel of Fig.~\ref{fig:post_photoz}, the 2PCF shape is significantly altered as photo-$z$ uncertainty increases, nearly erasing the BAO peak, which could potentially lead to a substantial loss of information. 
The right panel of Fig.~\ref{fig:post_photoz} shows that the constraints on $\om$ and $\sigma_8$ from clustering and weak lensing masses are only modestly affected in the case of $\sigma_{z0} = 0.005$, with a decrease in the FoM below 10\,\% compared to the case without photo-$z$ uncertainties, while a difference larger than 20\,\% is obtained when considering $\sigma_{z0} = 0.01$ . Such differences slightly increase when the full probe combination is considered, as a consequence of the tightening of the posteriors  (see Table~\ref{table:fom_phz}). More considerations about the impact of photo-$z$ uncertainties can be found in Appendix~\ref{appB}.

\begin{table}[t]
\centering           
\caption{Summary statistics for photo-$z$ impact on $\om$--$\sigma_8$ posteriors.}
\centering
\begin{tabular}{l c c c c}  
\hline
 & \multicolumn{2}{c}{ CL+M$_{\rm WL}$}& \multicolumn{2}{c}{NC+CL+M$_{\rm WL}$} \\
\hline  
 & $\Delta {\rm FoM} / {\rm FoM}$ & IoI & $\Delta {\rm FoM} / {\rm FoM}$ & IoI  \\
\hline
$\sigma_{z0}=0$     &   ...       & $+ 0.10$ & ...         & $+ 0.01$  \\
$\sigma_{z0}=0.005$ & $- 7$\,\%  & $+ 0.06$ & $- 10$\,\% & $+ 0.00$ \\
$\sigma_{z0}=0.01$  & $- 22$\,\% & $+ 0.11$ & $- 31$\,\% & $+ 0.01$ \\
\hline
\end{tabular}
\tablebib{FoM residuals are computed with respect to the true redshift case.  IoI is computed with respect to the fiducial cosmology and scaling relation parameters.}
\label{table:fom_phz} 
\end{table}

\subsubsection{Impact of secondary redshift distortions on clustering} \label{sec:results:clustering:zdist}
\begin{figure*}[t]
    \centering
    \includegraphics[width=0.45\textwidth]{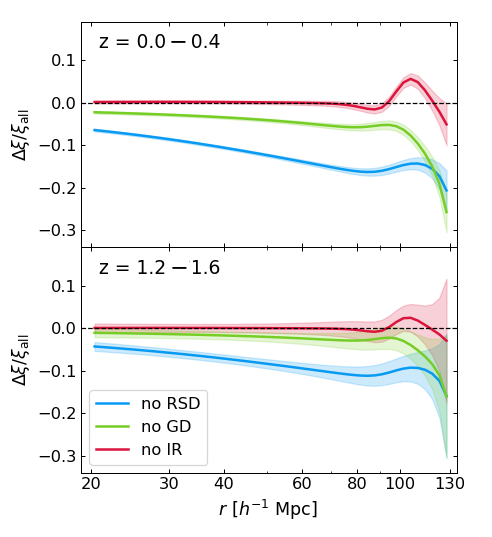}
    \includegraphics[width=0.484\textwidth]{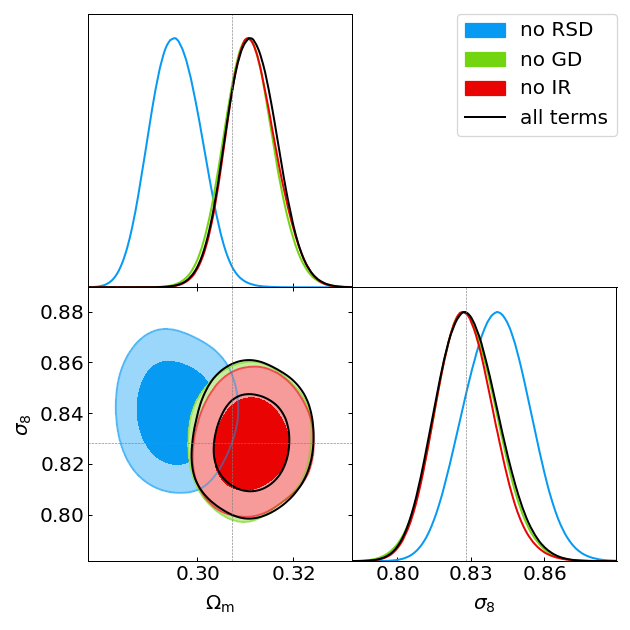}
    \caption{Systematic effects introduced by neglecting redshift distortions in 2PCF modelling. \textit{Left}: Residuals from the 2PCF with respect to the full model (i.e. photo-$z$, RSDs, IR resummation, and geometric distortion correction) for the three cases. The line colours are as follows: the model without RSDs, blue line; model without IR resummation, red line; and model without geometric distortion correction, green line). Shaded areas are given by the square root of the diagonal covariance divided by the number of mocks. \textit{Right}: Marginalised posterior distributions in the $\om$--$\sigma_8$ plane, with 68\% and 95\% confidence intervals, from the combination of clustering and weak lensing masses for the three cases represented in the left panel (same colour code) compared to the full redshift effects case (black empty contours). Dotted grey lines are the input cosmology of the catalogues.}
    \label{fig:2pcf_impacts}
\end{figure*}

As described in Sect.~\ref{sec:theory:2pcf}, the 2PCF is also affected by additional redshift distortions, namely RSDs, IR resummation, and geometric distortions. Although their effects are secondary to photo-$z$ uncertainties, neglecting them can still lead to significant biases in the cosmological posteriors. The left panel of Fig.~\ref{fig:2pcf_impacts} illustrates the distortions that occur when each of these effects is omitted from the 2PCF modelling but included in the measurement. In the case of RSDs, such a difference translates into a significant bias in the $\om$--$\sigma_8$ posteriors (right panel), with an ${\rm IoI} = 1.16$, and a decrease of $5\,\%$ in the FoM with respect to the full-modelling case (Table~\ref{table:fom_RSDs}). 
Instead, ignoring the IR resummation or the geometrical distortion results in almost no impact on the cosmological FoM nor in the position of the contours. 

When considering the full-probe analysis, the lack of RSDs modelling produces a shift in the posterior that is even larger (${\rm IoI} = 2.18$), due to the tightening of the contours. In all the other cases, the contours are well centred on the input cosmology. The difference in the FoM increases in all the cases, up to around $10\%$, indicating that neglecting each of the redshift effects may affect the final posteriors' amplitude. 

\begin{table}[t]
\centering           
\caption{Summary statistics for the impact of secondary redshift distortion on $\om$--$\sigma_8$ posteriors.}
\centering
\begin{tabular}{l c c c c}  
\hline
 & \multicolumn{2}{c}{ CL+M$_{\rm WL}$}& \multicolumn{2}{c}{NC+CL+M$_{\rm WL}$} \\
\hline  
 & $\Delta {\rm FoM} / {\rm FoM}$ & IoI & $\Delta {\rm FoM} / {\rm FoM}$ & IoI  \\
\hline
Full model   &   ...       & $+\,0.11$ & ...          & $+\,0.01$  \\
no RSDs       & $-\,5$\,\% & $+\,1.16$ & $+\,12$\,\% & $+ \,2.18$ \\
no IR        & $+\,3$\,\% & $+\,0.10$ & $+\,7$\,\%  & $+ \,0.00$ \\
no GD        & $+\,3$\,\% & $+\,0.07$ & $+\,9$\,\%  & $+ \,0.01$ \\
\hline
\end{tabular}
\tablebib{FoM residuals are computed with respect to the full model.  IoI is computed with respect to the fiducial cosmology and scaling relation parameters.}
\label{table:fom_RSDs} 
\end{table}

\begin{figure}[t]
    \centering
    \includegraphics[width=0.49\textwidth]{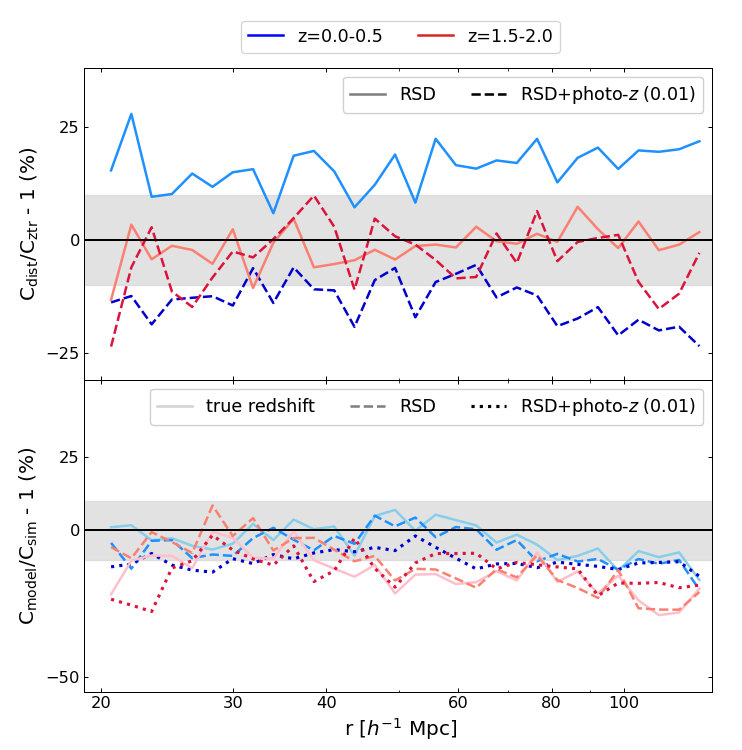}
    \caption{Residuals of the clustering covariance diagonal terms for the lowest (blue) and highest (red) redshift bins and $\lambda \in [20, 30]$. The grey area marks the 10\% region. \textit{Top}: Residuals of the distorted numerical covariance relative to the undistorted redshift case. Solid lines indicate the RSD case, while dashed lines include both RSDs and photo-$z$ effects ($\sigma_{z0} = 0.01$). \textit{Bottom}: Residuals between analytical and numerical covariances for different cases. Solid lines are for true redshift, dashed for RSDs, and dotted for both RSDs and photo-$z$ effects.}
    \label{fig:cl_cov_diff}
\end{figure}

The 2PCF covariance is largely dominated by shot-noise \citepalias[see the bottom panel of Fig. F.1 in][]{Fumagalli-EP27} and is therefore less affected by redshift distortions. As shown in the top panel of Fig.~\ref{fig:cl_cov_diff}, RSDs slightly increase the covariance between radial bins, while photo-$z$ uncertainties reduce it. Both effects are more pronounced at low redshift and diminish at higher redshift, where the cluster density decreases and shot-noise becomes more dominant.

The lower panel of Fig.~\ref{fig:cl_cov_diff} compares the residuals between model predictions (Eq.~\ref{eq:cov_cl}) and numerical results for different redshift distortions. The similarity of residuals across cases suggests that RSDs and photo-$z$ uncertainties do not alter the relationship between the model and numerical covariance. Thus, any differences between the model and the numerical matrix can be resolved by fitting the covariance parameters (defined in Sect.~\ref{sec:theory:2pcf}) as described in \citetalias{Fumagalli-EP27} and \citet{Fumagalli:2022plg}, maintaining their validity independent of redshift effects.

\subsubsection{Impact of radial scales in the clustering analysis} \label{sec:results:clustering:scales}
\begin{figure}[t]
    \centering
    \includegraphics[width=0.47\textwidth]{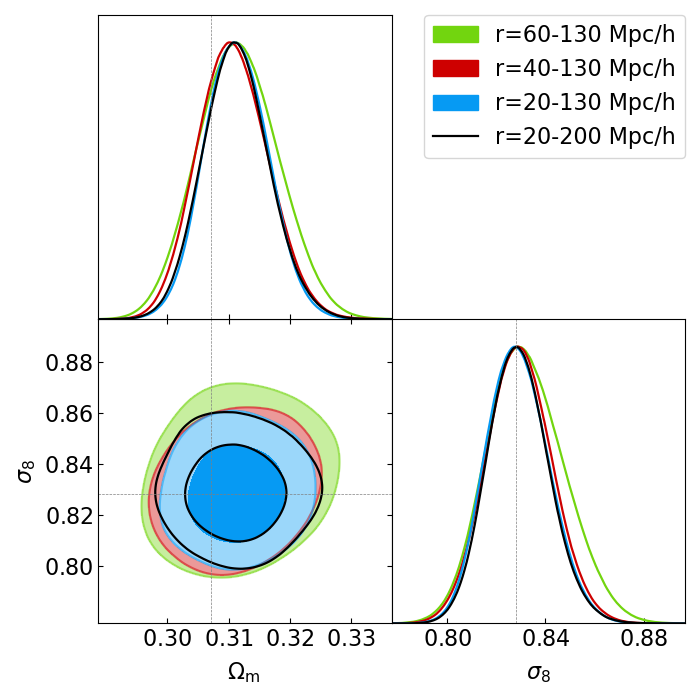}
    \caption{Marginalised posterior distributions with 68\% and 95\% confidence intervals from combined analysis of clustering and weak lensing masses for different clustering scales: $r=20$--$200\,h^{-1}\,$Mpc (in black), $r=20$--$130\,h^{-1}\,$Mpc (in blue), $r=40$--$130\,h^{-1}\,$Mpc (in red), and $r=60$--$130\,h^{-1}\,$Mpc (in green). Dotted grey lines are the input cosmology of the catalogues.}
    \label{fig:scales}
\end{figure}

In this section we investigate the impact of radial scales on the cosmological analysis of cluster clustering. As shown in \citet[][their Appendix C]{Fumagalli:2023yym}, the majority of the cosmological information comes from the region around the BAO peak (approximately $r = 60$--$130\,h^{-1}\,$Mpc), with small and large scales offering negligible contributions.  We repeat this analysis with a more detailed study, accounting for the properties of a \Euclid-like survey, such as extended redshift range and larger sample.  
By taking as reference the separation range $r = 20$--$130\,h^{-1}\,$Mpc, Fig.~\ref{fig:scales} and Table~\ref{table:fom_scales} show that increasing the lower limit to larger scales, such as  $r=40\,h^{-1}\,$Mpc or $r=60\,h^{-1}\,$Mpc, leads to a progressive broadening of the parameter constraints. This suggests that, for a survey such as \Euclid, scales between 20 and 60\,$h^{-1}\,$Mpc carry sizeable cosmological information.
On the other hand, we confirm that not much more information is extracted when extending the 2PCF analysis to larger scales, with a difference in the FoM of 3\% when increasing the upper limit to $200\,h^{-1}\,$Mpc. 
Similar results are obtained from the full-probe analysis.

\begin{table}[h]
\centering           
\caption{Summary statistics for the impact of different separation ranges on $\om$--$\sigma_8$ posteriors.}
\centering
\begin{tabular}{l c c c c}  
 & \multicolumn{2}{c}{ CL+M$_{\rm WL}$}& \multicolumn{2}{c}{NC+CL+M$_{\rm WL}$} \\
\hline  
 & $\Delta {\rm FoM} / {\rm FoM}$ & IoI & $\Delta {\rm FoM} / {\rm FoM}$ & IoI  \\
\hline
$r = 20\,$--$\,130$   &   ...        & $+\,0.10$ & ...          & $+\,0.01$  \\
$r = 40\,$--$\,130$   & $-\,14$\,\% & $+\,0.04$ & $-\,12$\,\% & $+\,0.01$ \\
$r = 60\,$--$\,130$   & $-\,37$\,\% & $+\,0.09$ & $-\,27$\,\% & $+\,0.01$ \\
$r = 20\,$--$\,200$   & $-\,3$\,\%  & $+\,0.09$ & $-\,2$\,\% & $+\,0.01$ \\
\hline
\end{tabular}
\tablebib{Radial separations in the first column are in $h^{-1}\,$Mpc units. Figure of merit residuals are computed with respect to the separation range $r=20\,$--$\,130\,h^{-1}\,$Mpc.  IoI is computed with respect to the fiducial cosmology and scaling relation parameters.}
\label{table:fom_scales} 
\end{table}

\subsection{Number count validation} \label{sec:results:numbercounts}
\begin{figure}[t]
    \centering
    \includegraphics[width=0.49\textwidth]{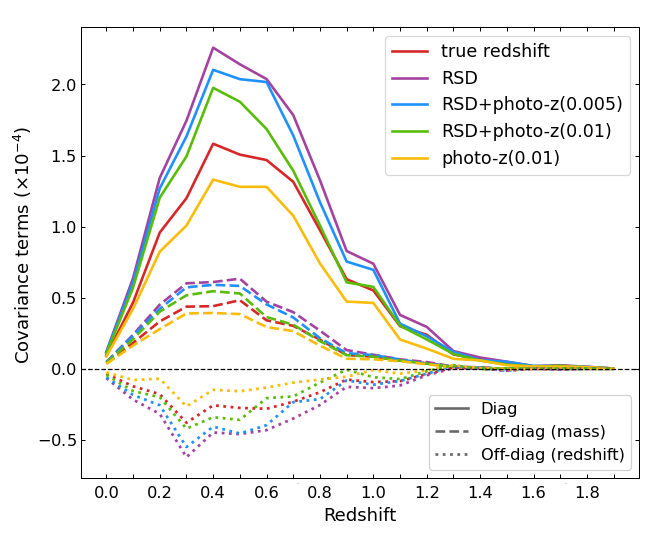}
    \caption{Number count sample covariance terms from numerical simulations for different redshift settings: undistorted (red lines), RSDs (purple lines), RSDs and small photo-$z$ uncertainty ($\sigma_{z0}=0.005$, blue lines), RSDs and large photo-$z$ uncertainty ($\sigma_{z0}=0.01$, green lines), and redshift with only large photo-$z$ uncertainty ($\sigma_{z0}=0.01$, yellow lines). Solid lines represent the diagonal sample variance, dashed lines are the sample covariance between two adjacent mass bins (first and second) and same redshift bin, and dotted lines are the sample covariance between adjacent redshift bins and same mass bin (first one).}
    \label{fig:nc_cov_z}
\end{figure}

This section examines the effects of RSDs and photo-$z$ redshift errors on cluster number counts and its covariance. The presence of redshift effects, especially photo-$z$ uncertainties, shifts redshifts from the centre of the distribution to the edges, potentially altering the number of objects in each redshift bin. However, this does not affect the overall shape of the cluster abundance distribution, leaving the shot-noise term unchanged.
In contrast, the sample variance signal, which is determined by the clustering signal integrated over the redshift bins (Eq.~\ref{eq:nc_sample_cov}), reflects the distortions in the power spectrum caused by RSDs and photo-$z$ uncertainties. As shown in Fig.~\ref{fig:nc_cov_z}, RSDs tend to amplify sample variance, whereas photo-$z$ uncertainties reduce it. The increase due to RSDs arises because large-scale velocity flows, which drive RSDs on linear scales, are predominantly sourced by long-wavelength (small-$k$) density fluctuations -- the same modes that contribute to sample variance. As a result, RSDs enhance the impact of these fluctuations, leading to a higher sample variance.
Conversely, photo-$z$ uncertainties blur the positions of clusters along the line of sight, effectively smoothing the density field in redshift space. This reduces the contrast of density fluctuations, leading to a decrease in sample variance.

\begin{figure}[t]
    \centering
    \includegraphics[width=0.49\textwidth]{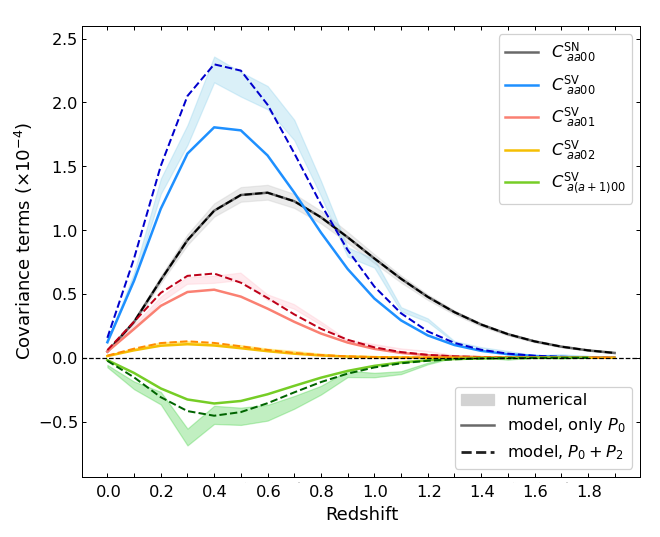}
    \caption{Number count covariance terms:  numerical matrix (shaded areas, representing the 1$\sigma$ region), analytical model with power spectrum monopole (solid lines), and model with effective power spectrum (see Eq.~\ref{eq:eff_pk}, dashed lines). The colour-coded terms represent different components: Grey is for diagonal shot-noise; blue is for diagonal sample variance; red and yellow are for the first and second off-diagonal sample covariance between mass bins, respectively; and green is for the first off-diagonal sample covariance in redshift bins. }
    \label{fig:nc_cov_RSD}
\end{figure}

Incorporating in the model the RSDs and photo-$z$ effects solely through the monopole of the power spectrum proves insufficient to recover the numerical results, as shown in Fig.~\ref{fig:nc_cov_RSD}. On the other hand, including the correct modelling of the anisotropic power spectrum in Eq.~\eqref{eq:nc_sample_cov} would break many symmetries and lead to a computationally expensive calculation.
Dashed lines in Fig.~\ref{fig:nc_cov_RSD} show that an effective power spectrum given simply by the sum of the power spectrum multipoles, i.e.
\begin{equation}\label{eq:eff_pk}
    P_{\text{eff}} (k) = \sum_{\ell=0,2,4} P_\ell(k)\,,
\end{equation}
effectively captures the sample covariance in a computationally efficient manner, balancing accuracy with practical feasibility. Furthermore, since the hexadecapole ($\ell=4$) is found to be negligible, the sum can be restricted to the monopole and quadrupole terms.

\begin{figure}[t]
    \centering
    \includegraphics[width=0.47\textwidth]{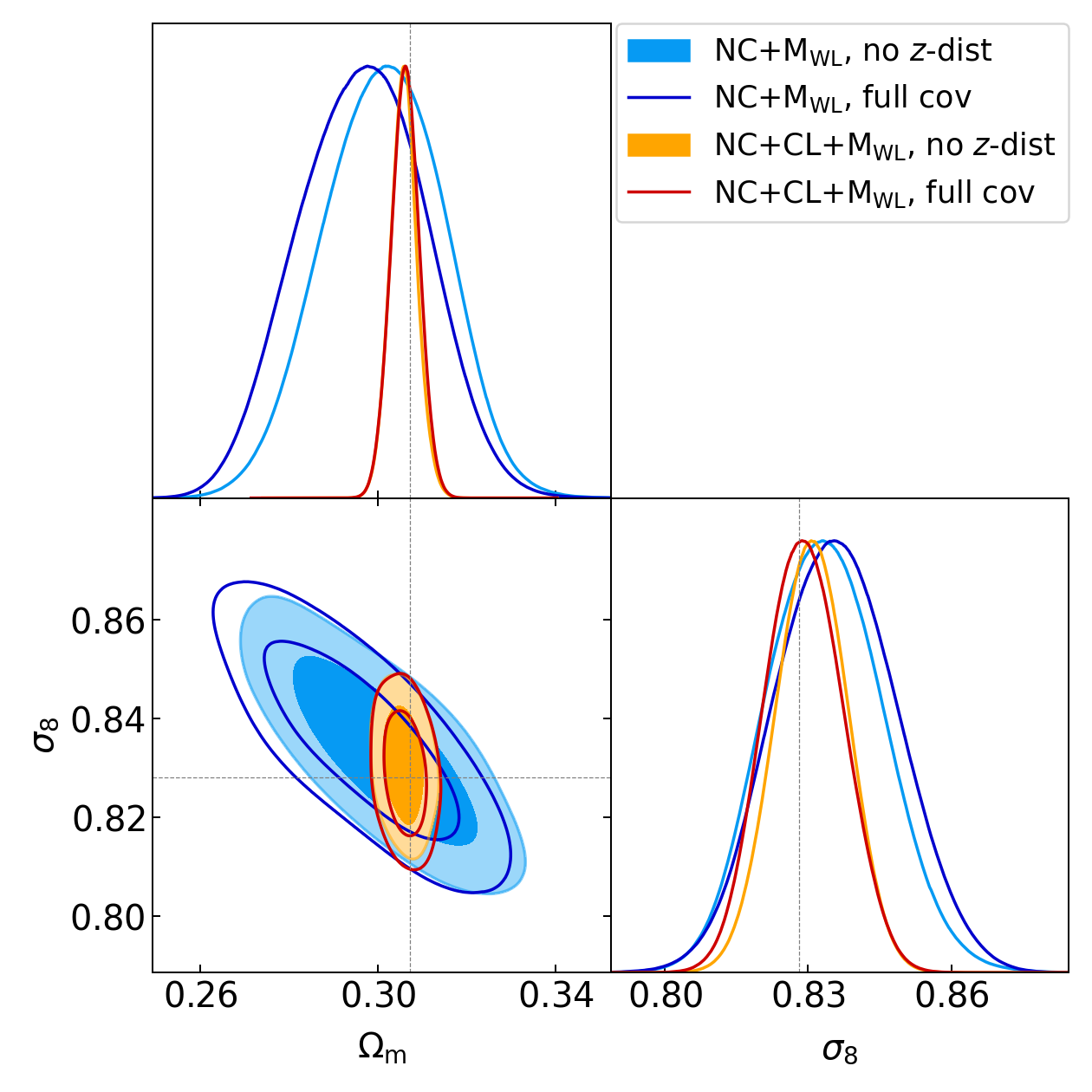}
    \caption{Marginalised posterior distributions with 68\% and 95\% confidence intervals from number counts (blue and cyan contours) and number counts with clustering (red and orange contours), with (empty contours) and without (filled contours) the proper modelling of redshift distortions in the number count covariance. Dotted grey lines are the input cosmology of the catalogues.}
    \label{fig:post_nccov}
\end{figure}

We find that neglecting photo-$z$ and RSD effects in the covariance significantly impacts cosmological constraints, as shown in Fig.~\ref{fig:post_nccov}. In the photo-$z$ case with $\sigma_{z0} = 0.01$, this leads to an underestimation of the posterior amplitude by about 20\%. The effect is expected to be even larger for smaller photo-$z$ uncertainties, as the difference in sample variance increases. A similar trend is observed in the full-probe analysis, where neglecting RSDs and photo-$z$ in the number count covariance degrades the FoM by 10\% compared to proper covariance modelling.

\section{Conclusions} \label{sec:conclusions}
In this paper, we have shown a framework to extract cosmological information from the combination of number counts and clustering of galaxy clusters in a survey with a mass selection, sky coverage, and depth similar to those expected from \Euclid.
Building on the models from \citetalias{Fumagalli21} and \citetalias{Fumagalli-EP27}, we extended the methodology to explicitly include the effect of RSDs as well as observational uncertainties, such as errors in photo-$z$ and in the mass-richness relation. We assessed the impact of including these effects on number counts, the monopole of the 2PCF, and their respective covariance matrices.

Our validation was conducted using 1000 \Euclid-like light cones generated with the \texttt{PINOCCHIO} code, where we introduced observational effects such as photometric and richness estimate errors. To quantify the influence of these distortions, we performed a likelihood analysis across different configurations and evaluated their impact on cosmological posteriors. The analysis focused on constraining $\om$, $\sigma_8$, and the four scaling relation parameters (Eqs.~\ref{eq:mor_mean} and~\ref{eq:mor_var}). 

The key findings of our analysis can be summarised as follows.
\begin{itemize}
    \item We demonstrated in Sect.~\ref{sec:results:crosscov} that numerical covariances show no significant cross-correlation between number counts and 2PCF, with a Pearson correlation factor consistent with zero. The two statistics are therefore independent. Such a result is in agreement with \citet{Mana:2013qba} and \cite{Fumagalli:2023yym}.
    \item We find that the full-probe analysis is the most powerful combination for extracting cosmological information and the less affected by prior-volume effects.
    \item We confirmed in Sect.~\ref{sec:results:cosmocov} that the cosmology-dependence of the clustering covariance brings useful information, increasing the constraining power of clustering by about 40\%. Such an improvement disappears when number counts are combined, which is in line with the result that number counts already encapsulate additional information from the clustering covariance. Similarly, evaluating the covariance at a wrong cosmology has a significant impact on the clustering-only case, an effect that decreases when adding number counts.
    \item The primary effect of redshift effects on the 2PCF arises from photo-$z$ uncertainties. We showed in Sect.~\ref{sec:results:clustering:phz} that an optimistic estimate for photo-$z$ uncertainties of $\sigma_{z0} = 0.005$ results in a small impact on parameter constraints, reducing the cosmological FoM by less than 10\%. However, for a more conservative value of $\sigma_{z0} = 0.01$, the posteriors broaden significantly, with a decrease in the FoM of 20--30\%.
    \item We showed in Sect.~\ref{sec:results:clustering:zdist} that secondary redshift distortions -- namely, RSDs, IR resummation, and geometric distortions -- have an almost negligible impact ($\lesssim 5\%$) on the FoM when clustering is considered alone. This impact slightly increases to around 10\% when combining all probes in the analysis. In addition, the absence of proper RSD modelling leads to a non-negligible shift in the contours  (about $2 \sigma$).
    \item Although most of the cosmological information in the linear regime is contained at large scales ($r = 60$--$130\,h^{-1}\,$Mpc), we showed in Sect.~\ref{sec:results:clustering:scales} that smaller scales also contribute to increases in the constraining power of clustering. In contrast, expanding the analysis to scales larger than $130\,h^{-1}\,$Mpc, does not bring any significant improvement.
    \item While RSDs and photo-$z$ uncertainties have no direct impact on the number counts themselves, we found that they do influence the number count covariance (Sect.~\ref{sec:results:numbercounts}). Specifically, RSDs increase the sample variance, whereas photo-$z$ uncertainties decrease it. Neglecting these effects in the covariance modelling can lead to a 10--20\% underestimation of the posterior amplitude. We propose an effective model that accurately incorporates these effects into the covariance without adding unnecessary computational burden to the model.
\end{itemize}

Our work underscores the benefits of combining cluster clustering with traditional number counts. 
First, we note that the use of a wide redshift range allows the amount of extracted information to be greatly increased. This is observed, in particular, through the combination of number counts and clustering. In \citet{Fumagalli:2023yym}, where a single low-redshift bin was analysed, the combination of these two probes failed to constrain $\sigma_8$. In contrast, with multiple bins spanning a wider redshift range, the constraints become much tighter and almost competitive with the combination of number counts and weak lensing mass, highlighting how the redshift dependence of the halo mass function and halo bias provides substantial additional information. 

Second, in addition to confirming that the combination of the three probes provides the most stringent constraints -- with an FoM improvement of more than $300\%$ with respect to the standard combination of number counts and weak lensing masses -- we also find that these results are better centred around the input cosmology. This indicates that the addition of clustering not only enhances the extraction of cosmological information but also mitigates the impact of systematics, leading to more precise and more accurate constraints. The absence of significant correlations between number counts and clustering simplifies the analysis, further optimising the information gain. 
Moreover, this combination is less sensitive to the cosmological dependence of the covariances, as the cosmological dependence of the observables saturates the information that can be extracted from the sample. 

However, our findings also highlight the need for caution when performing full-probe analyses, as narrower parameter constraints make them more sensitive to systematic effects. Some uncertainties that are negligible when considering clustering or number counts independently may become significant when both probes are combined, stressing the importance of carefully modelling systematic uncertainties. 

It is important to emphasise that the results of this work do not fully reflect the complexity of the future \Euclid cluster-cosmology analysis. The presented analysis is rather conservative, assuming mass calibration at the 1\% level out to redshift $z=2$ and neglecting potential model systematics, such as uncertainties in the $P(\lob|M)$ relation at high redshift. In particular, the impact of projection effects on cluster number counts, clustering, and their combination remains to be quantified. In addition, some key elements still need to be explored. Firstly, current analyses neglect the cross-correlation between weak lensing mass and both counts and clustering. While this assumption is supported by reasonable arguments, it needs to be tested. Additionally, the impacts of the survey mask and the presence of different homogeneous regions on the covariances of clustering and counts still require investigation. Finally, the clustering analysis is not fully explored due to the absence of higher-order multipoles. Given that photo-$z$ uncertainties induce a non-negligible signal in these components, their inclusion could significantly reduce the final uncertainties on the parameters.
Nevertheless, these results provide valuable insight into the influence of observational and modelling systematics on cosmological constraints from upcoming \Euclid data, highlighting the necessity of accounting for effects that were neglected in previous studies.

\begin{acknowledgements}
AF acknowledges support by the Excellence Cluster ORIGINS, which is funded by the Deutsche Forschungsgemeinschaft (DFG, German Research Foundation) under Germany’s Excellence Strategy - EXC-2094 - 390783311 and from the Ludwig-Maximilians-Universit\"at in Munich. This paper is supported by the Agenzia Spaziale Italiana (ASI) under - Euclid-FASE D  Attivita' scientifica per la missione - Accordo attuativo ASI-INAF n. 2018-23-HH.0, by 
the National Recovery and Resilience Plan (NRRP), Mission 4,
Component 2, Investment 1.1, Call for tender No. 1409 published on
14.9.2022 by the Italian Ministry of University and Research (MUR),
funded by the European Union – NextGenerationEU– Project Title
"Space-based cosmology with Euclid: the role of High-Performance
Computing" – CUP J53D23019100001 - Grant Assignment Decree No. 962
adopted on 30/06/2023 by the Italian Ministry of Ministry of
University and Research (MUR);
by the Italian Research centre on High-Performance Computing Big Data and Quantum Computing (ICSC), a project funded by European Union - NextGenerationEU - and National Recovery and Resilience Plan (NRRP) - Mission 4 Component 2, by the INFN INDARK PD51 grant, and by the PRIN 2022 project EMC2 - Euclid Mission Cluster Cosmology: unlock the full cosmological utility of the Euclid photometric cluster catalogue (code no. J53D23001620006). MR and FM acknowledge the financial contribution from the PRIN-MUR 2022 20227RNLY3 grant “The concordance cosmological model: stress-tests with galaxy clusters” supported by Next Generation EU and from the grant ASI n. 2024-10-HH.0 “Attività scientifiche per la missione Euclid – fase E”. ET acknowledges support by STFC through Imperial College Astrophysics Consolidated Grant ST/W000989/1.

 \AckEC
\end{acknowledgements}

\bibliographystyle{aa} 
\bibliography{biblio}

\begin{appendix} 
\section{Impact of redshift photometric bias} \label{appA}
\begin{figure}[h]
    \centering
    \includegraphics[width=0.49\textwidth]{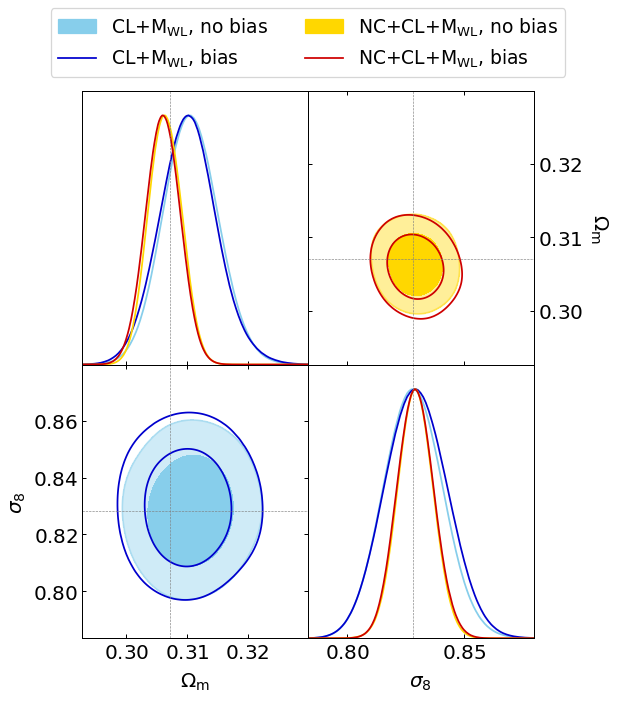}
    \caption{Marginalised posterior distributions in the $\om$--$\sigma_8$ plane, with 68\% and 95\% confidence intervals, with (filled contours), and without (empty contours), photo-$z$ bias. The results correspond to the optimistic photo-$z$ uncertainty amplitude. \textit{Lower left}: Clustering and weak lensing masses alone. \textit{Upper right}: Full-probe combination. }
    \label{fig:phz_shift}
\end{figure}

We check here the impact of the photo-$z$ bias on cosmological constraints. For this test, we modified the redshift selection function $P(\zob\,|\,\ztr, \Delta \lob_i)$ in Eq.~\eqref{eq:obs_vol} by adding a shift to the true redshift of $\Delta \ztr = 0.002$. Figure~\ref{fig:phz_shift} shows that, even in the optimistic photo-$z$ case ($\sigma_0 = 0.005$), such a redshift bias does not impact the cosmological constraints significantly. Both the clustering and weak lensing masses and the full-probe combinations are characterised by no difference in the IoI between the bias and no-bias cases, as well as by a $\Delta$FoM consistent with zero.
We therefore do not add any bias to the observed redshift in out simulations (see Sect.~\ref{sec:data:sims}), as the effect is negligible.

\section{Impact of non-linearities} \label{appB}
\begin{figure}[t]
    \centering
    \includegraphics[width=0.5\textwidth]{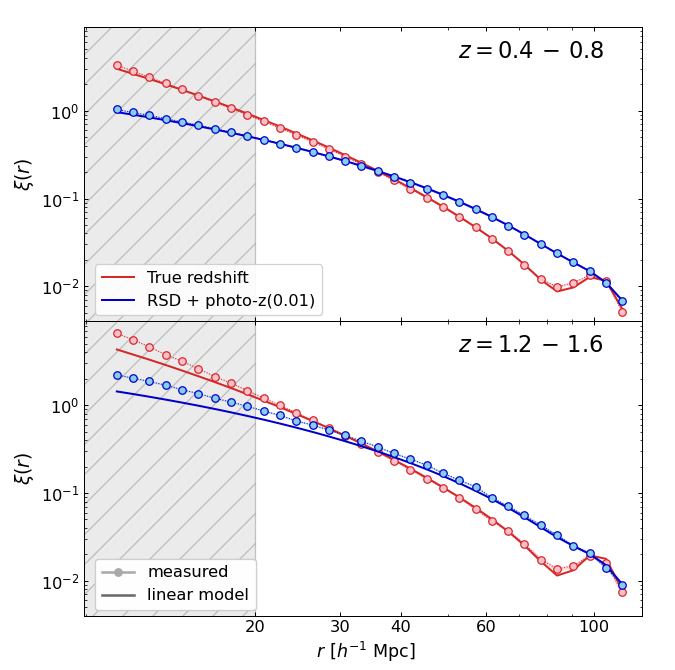}
    \caption{Two-point correlation function without (red) and with (blue) photo-$z$ uncertainty ($\sigma_{z0}=0.01$). Solid lines are the linear model, and circles are the measurements, averaged over the 1000 mocks. The standard error on the mean is included but too small to be noticeable. The grey area indicates scales not included in the cosmological analysis. \textit{Top}: Low-redshift bin. \textit{Bottom}: High-redshift bin.}
    \label{fig:2pcf_photoz}
\end{figure}

When including the effect of photo-$z$ uncertainties, we notice a discrepancy between our model and the measured 2PCF. Although the model outlined in Sect.~\ref{sec:theory:2pcf} accounts for photo-$z$-induced distortions in the 2PCF shape, Fig.~\ref{fig:2pcf_photoz} highlights a scale-dependent mismatch between the predicted and observed 2PCF when photo-$z$ uncertainties are included. The effect, which  increases with redshift, becomes significant for $z \gtrsim 1$. A similar discrepancy characterises the undistorted 2PCF, but is confined to the non-linear regime ($r \lesssim 20\,h^{-1}\,$Mpc). Photo-$z$ uncertainties mix the power spectrum modes parallel to the line of sight, propagating non-linearities to larger scale and making the linear theory inadequate for describing clustering up to scales of approximately $60\,h^{-1}\,$Mpc at high redshift. Non-linear distortions also explain the redshift dependence of this mismatch: Objects of fixed mass are characterised by a higher bias and, therefore, increasing non-linearity at higher redshifts.

\begin{figure}[h]
    \centering
    \includegraphics[width=0.5\textwidth]{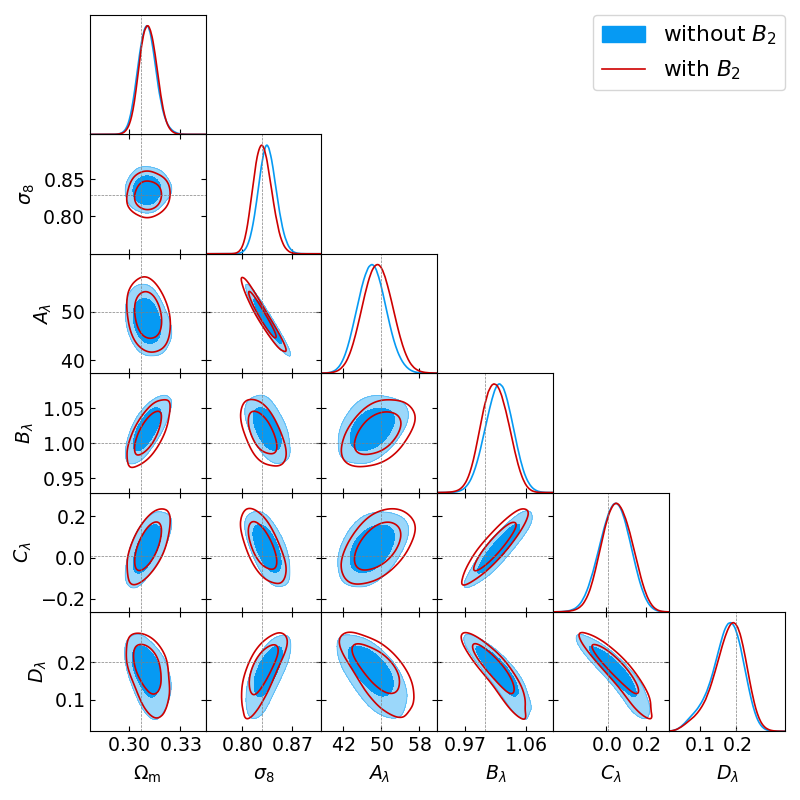}
    \caption{Parameter posteriors with 68\% and 95\% confidence intervals, from clustering and weak lensing mass analysis, with (blue contours) and without (empty red contours) the bias correction factor $B_2$.}
    \label{fig:post_b2}
\end{figure}

Addressing this requires a non-linear model for the halo bias, which is beyond the scope of this paper. Instead, we introduce a correction factor, defined as
\begin{equation} \label{eq:bcorr}
    B_2 = \frac{\hat{\xi}_{\rm obs}}{\xi_{\rm fid}},
\end{equation}
where $\hat{\xi}_{\rm obs}$ is the measured 2PCF, $\xi_{\rm fid}$ is the prediction at the fiducial cosmology, and $B_2$ is a scale-dependent correction factor for each redshift and richness bin. This factor is included in the likelihood analysis to enhance the model accuracy. However, it serves primarily as a `safety net' and is not strictly essential: as shown in Fig.~\ref{fig:post_b2}, this discrepancy does not lead to significant differences in the cosmological posteriors. 

With low-redshift and large scales dominating the clustering constraining power, the high-redshift, small-scale regime contributes only marginally to the cosmological constraints.

\begin{figure}[h]
    \centering
    \includegraphics[width=0.5\textwidth]{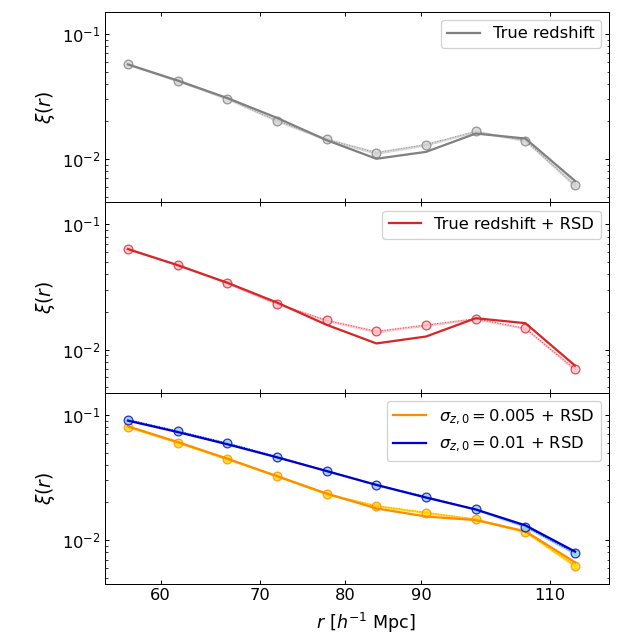}
    \caption{Impact of non-linearities around the BAO peak. Model (solid lines) and measured (circles) 2PCF with true redshift (top), true redshift with RSDs (centre), and observed redshift with RSDs and photo-$z$ (top).}
    \label{fig:2pcf_RSD}
\end{figure}

We also notice a small discrepancy between model and measurements around the BAO peak, when true redshifts are considered (top panel of Fig.~\ref{fig:2pcf_RSD}). Such a difference increases when RSDs are included, as shown in the central panel. Again, the cause can be attributed to the failure of linear theory in IR re-summation and RSDs modelling. However, as can be seen from the bottom panel, in the presence of photo-$z$ uncertainties this effect is completely washed out.

\end{appendix}
\end{document}